\documentclass[twocolumn]{aastex631}

\usepackage{float}      
\usepackage{booktabs}    
\usepackage{siunitx}     
\usepackage{multirow}    
\usepackage{amsmath}     
\usepackage{graphicx}   
\usepackage{makecell}   
\usepackage{array}     
\usepackage{booktabs}
\usepackage{fancyhdr}
\usepackage{hyperref} 

\shorttitle{Mock Catalog of lensed GWs}

\shortauthors{Li et al.}

\begin{document}

\title{Mock Catalogs of Strongly Lensed Gravitational Waves via A Halo Model Approach with Ground-based Detectors}

\correspondingauthor{Kai Liao}
\email{liaokai@whu.edu.cn}

\author{Youkai Li}
\affiliation{School of Physics and Technology, Wuhan University, Wuhan 430072, China;}

\author{Kai Liao*}
\affiliation{School of Physics and Technology, Wuhan University, Wuhan 430072, China;}

\author{Mingqi Sun}
\affiliation{School of Physics and Technology, Wuhan University, Wuhan 430072, China;}

\author{Lilan Yang}
\affiliation{School of Physics and Technology, Wuhan University, Wuhan 430072, China;}

\author{Xuheng Ding}
\affiliation{School of Physics and Technology, Wuhan University, Wuhan 430072, China;}

\author{Marek Biesiada}
\affiliation{National Centre for Nuclear Research, Pasteura 7, PL-02-093 Warsaw, Poland;}
\author{Tonghua Liu}
\affiliation{School of Physics and Optoelectronic, Yangtze University, Jingzhou 434023, China;}

\begin{abstract}
As plans for the construction of third-generation gravitational wave (GW) detectors advance, research into strongly lensed GWs has become increasingly critical. It is anticipated that hundreds of multi-image lensed GWs will be detected annually. We present a comprehensive suite of lensed GW mock catalog derived from a composite lens mass model incorporating dark matter halos, galaxies, and subhalos. We analyze three source populations with four detector network configurations considering the earth rotation. Our simulations encompass not only conventional doublets and quadruplets but also subhalo-lensed events, highly magnified systems, and complete three or five image systems with a detectable central image, a feature distinct from optical lensing. For the joint ET+CE network, we forecast an annual detection rate of approximately 400 doublets and 36 quadruplets. Notably, this population includes roughly 107 events lensed by subhalos and 20 complete systems with detectable central images. Furthermore, we analyze high-magnification events ($\mu > 3$), predicting approximately 360 such cases. Under a more relaxed selection criterion that requires only at least one lensed signal to exceed the detection threshold, we estimate a total of approximately 617 lensed events. We also investigate the impact of variations in lens mass models and stellar evolution models on event rates, as well as the distributions of SNR pairs and time delays. These results establish a more physically grounded statistical prior for the future identification and authentication of lensed GW signals. The Gravitational Waves-Lensing Mock Catalog (GW-LMC) have been made publicly available.
\end{abstract}

\section{Introduction}
With the successful conclusion of the first half of the fourth observing run (O4a) by the Laser Interferometer GW Observatory (LIGO; \citet{Aasi_2015}), Virgo (\citealt{Acernese_2015}), and  {the Kamioka Gravitational Wave Detector (KAGRA; \citealt{Abbott2020})}, we have secured an increasingly rich population sample of GWs. On August 26, 2025, the LVK Collaboration released the latest GW transient catalog, GWTC-4.0, which includes 128 new significant candidates confirmed during the O4a observing run (May 2023 to January 2024) \citep{2025arXiv250818082T}. These data constitute the primary dataset for current lensing searches.

Analogous to electromagnetic wave, GW are subject to gravitational lensing, a phenomenon wherein massive objects deflect their propagation paths \citep{pub.1007554514,osti_1155593,10.1093/mnras/sty2145}. Strong lensing can generate multiple images of a single source, characterized by distinct differences in arrival time, amplitude, and phase \citep{dai2017waveformsgravitationallylensedgravitational}. The observational signatures depend critically on the ratio of the lens characteristic scale to the GW wavelength. In the geometric optics regime (short wavelength), lensing magnifies signals, enabling the detection of lower-mass or higher-redshift binaries \citep{PhysRevD.95.044011}. Conversely, in the wave optics regime--where the lens size is comparable to the wavelength--diffraction effects emerge, inducing significant distortions in the GW waveforms \citep{Takahashi_2003}.

The identification of lensed GW signals offers a novel probe for fundamental physics and cosmology. In fundamental physics, repeated signals allow for stringent tests of General Relativity \citep{PhysRevD.102.124048}. {A} lensed gravitational signal registered together with an electromagnetic counterpart would allow to measure the speed of GW directly \citep{PRL2017,2017PhRvL.118i1101C}. Cosmologically, the millisecond-level precision of GW arrival times provides an ideal, independent tool for measuring the Hubble constant ($H_0$) and addressing the Hubble tension \citep{Liao2017}. Furthermore, detecting lensed GWs aids in probing large-scale structures and the nature of dark matter by breaking the mass-sheet degeneracy \citep{PhysRevD.104.023503,PhysRevD.108.103529}. Lensed GW also create an opportunity to measure dark matter self-interaction cross section \citep{DMviscosity1,DMviscosity2}. Finally, in the era of multi-messenger astronomy, lensed GWs facilitate the study of binary merger environments and can provide early warnings for electromagnetic follow-up \citep{Magare_2023}. For more on lensed GWs, we refer to the review papers \citep{2019RPPh...82l6901O,2022ChPhL..39k9801L}.

Only a minute fraction of GW events is expected to undergo strong lensing\citep{2014JCAP,Xuheng2015,PhysRevD.97.023012,10.1093/mnras/sty411,10.1093/mnras/stab3298,Wierda_2021,Xu_2022}. However, the expanding catalog size leads to a quadratic growth in candidate pairs, causing a sharp rise in false alarm rates due to chance coincidences \citep{_al_kan_2023}. Fortunately, recent studies suggest that with precise time-delay models, the bias from selection effects is neutralized, preserving detection capabilities even as catalogs grow \citep{hannuksela2025stronggravitationalwavelensingposterior,Xu_2022}. Current identification strategies primarily employ: parameter overlapping \citep{2018arXiv180707062H,Barsode_2025}, joint parameter estimation to evaluate Bayes factors for multiple images \citep{Liu_2021,PhysRevD.107.123015,10.1093/mnras/stab1991}, search for characteristic Morse phase shifts or diffractive distortions in single signals \citep{Takahashi_2003,Ezquiaga2021,Janquart_2021}, an interference-based method \citep{2025NatAs...9..916S}, with host information \citep{2025arXiv251210344L}. To address computational demands, these methods are increasingly augmented by rapid post-processing techniques \citep{Janquart_2023} and cutting-edge AI applications, such as neural posterior estimation and CNN-based classification \citep{wong2023fastgravitationalwaveparameter,Magare_2024,2025arXiv250819311L,2025arXiv250904538L,2026ApJS..283...31L}.

For the O4a dataset, a multi-dimensional search strategy was adopted to comprehensively capture potential lensing signatures \citep{theligoscientificcollaboration2026gwtc40searchesgravitationalwavelensing}. In contrast to the galaxy-scale strong lensing discussed in subsequent sections, which typically produces multiple discrete images, the search for single signals focuses on lensing effects at smaller mass scales (e.g., microlensing or millilensing). Some recent studies have conducted a preliminary analysis of the GW231123 event, revealing that there are significant differences in the measurement of source characteristics across different waveform models of the event. These differences cannot be reliably reproduced through standard zero-noise injection-recovery simulations with incomplete waveform physics, as documented in prior work on heavy, rapidly spinning  (BBH) systems \citep{2026PhRvD.113h3001X,2025arXiv251220060J}. These discrepancies may arise from unconsidered effects in the data—including but not limited to microlensing—alongside other plausible explanations such as orbital eccentricity, violation of the dispersion relationship, or residual waveform systematics \citep{goyal2025universegw231123magnifieddiffracted,chan2026discoveringgravitationalwaveformdistortions,hu2025gw231123overlappinggravitationalwave,shan2025gw231123casebinarymicrolensing,2025arXiv251220060J,2026arXiv260408179W,2026arXiv260319020T}.
As one of the most massive BBH systems in O4a (with a total mass of $190$--$265\,M_\odot$), GW231123 remains a candidate of particular interest. However, definitive confirmation of lensing signatures is currently hindered by waveform uncertainties and the complex noise properties of the detectors \citep{ray2025gw231123extremespinsmicroglitches,bini2026impactwaveformsystematicsgaussian}, alongside the concern that the expanded parameter space of the lens model may simply be fitting unmodeled features within the data. Furthermore, other proposed explanations for the observed signal distortions, such as overlapping independent signals \citep{hu2025gw231123overlappinggravitationalwave} or orbital eccentricity—consistent with recent analyses of GWTC-4.0 high-mass events \citep{2026PhRvD.113h3001X}-cannot yet be ruled out. These alternative scenarios collectively contribute to the difficulty in reaching a definitive conclusion on the lensing hypothesis for this event.

This paper presents a multi-scale compound lens model \citep{Abe_2025} based on the Halo Model \citep[see e.g.,][]{Asgari_2023}, designed to transcend the limitations of traditional galaxy-scale-only predictions (e.g., OM10; \citealt{Oguri_Marshall_2010}). By superimposing dark matter halos with baryonic components, this model unifies lensing predictions across galaxy, group, and cluster scales, and explicitly incorporates the contributions of subhalos \citep{Oguri_Takahashi_2020} and satellite galaxies which are important for lensed GWs which have high-resolution in time domain for multiple images .

This work aims to present a comprehensive and complete GW lensing mock catalog based on a composite lens model. The structure of the paper is as follows. In Section~2, we outline the distributions adopted for the merger rates, masses, and spins of the GW events, employing physically motivated distribution models. Specifically, for BBH mergers, we utilize the ``Broken Power Law + 2 Peaks'' mass model derived from  \citet{theligoscientificcollaboration2025gwtc40populationpropertiesmerging}, which offers a more realistic physical representation than the power-law models used in previous studies \citep{10.1093/mnras/stab3298}. Section~3 briefly introduces the composite lens model employed in this study. This model establishes a unified framework integrating dark matter halos, galactic potentials, subhalos, and external shear, facilitating a smooth transition from galaxy-scale to group- and cluster-scale lensing \citep{Abe_2025}. Section~4 describes the sampling methodology for GW source parameters and the generation of lensed GW results based on the \texttt{SL-Hammocks} code. We also detail the selection criteria applied to the results. Utilizing this model, we derive detailed parameters for lensed GW signals, including time delays, magnification factors, convergence ($\kappa$), and shear ($\gamma$). Section~5 details the mock catalog, which covers predictions for all categories of lensed GW events (BBH, NSBH, BNS) across four detector configurations: the A+ upgrade (PlusNetwork) \citep{Barsotti2018}, CE \citep{evans2021horizonstudycosmicexplorer}, ET \citep{Maggiore_2020}, and the combined ET+CE network.

We expand upon the work of \citet{10.1093/mnras/stab3298}, which focused solely on double-image events, by discussing standard multiple-image systems (doubles and quads), events lensed by dark matter subhalos, and crucially, ``central image'' events. According to the odd-number theorem \citep{1981ApJ...244L...1B}, multiple-image systems physically comprise 3 or 5 images. However, one image typically forms very close to the lens center and undergoes strong demagnification; rendered invisible in optical astronomy by the glare of the central galaxy \citep{Winn2004}, these systems are conventionally classified as doubles or quads. In the context of GWs, however, these central images remain detectable if the SNR is sufficiently high. Furthermore, we provide priors on SNR ratios and time delay distributions to support lens identification algorithms.These results establish a standardized benchmark for evaluating lensed GW signals, thereby improving algorithmic efficiency and precision. Finally, all results and the GW-LMC (Gravitational Waves-Lensing Mock Catalog) are publicly available on GitHub\footnote{\url{https://github.com/LensedGW/GW-LMC}} and have been archived on Zenodo\footnote{\url{https://zenodo.org/records/19212271}} \citep{li_youkai_2026_19212271}.

\section{Source Model}
The intensity of gravitational lensing effects depends not only on the alignment geometry of the source, the lens, and the observer but also on the lens mass and the wavelength of the GW. Variations in the source population distribution directly influence the predicted rates of lensed GW events; conversely, the construction of the source parameter model governs the intrinsic strength of the GW signals. Therefore, in this section, we provide a detailed description of the GW source models adopted in this work.

\subsection{Redshift Distribution}
When assessing the susceptibility of GW events to strong lensing, the GW event rate is the primary quantity of interest. Using the intrinsic merger rate formalism for compact binary coalescences (CBCs), we derive the observed event rate distribution as a function of redshift:
\begin{equation}
    \frac{\mathrm{d}\dot{N}}{\mathrm{d}z_s}=4\pi\left(\frac{c}{H_0}\right)^3\frac{\dot{n}_0(z_s)}{1+z_s}\frac{\tilde{r}^2(z_s)}{E(z_s)},
    \label{eq:merger_rate}
\end{equation}
where $\dot{n}_0(z_s)$ represents the intrinsic merger rate density at the source redshift $z_s$; $\tilde{r}(z_s)$ denotes the dimensionless comoving distance, defined as $\tilde{r}(z_s) = \int_0^{z_s} \frac{\mathrm{d}z'}{E(z')}$; and $E(z_s)$ is the dimensionless expansion rate of the Universe at redshift $z_s$.
Although \citet{theligoscientificcollaboration2025gwtc40populationpropertiesmerging} provides the latest data from the O4a observing run, current observational sensitivities restrict the detection horizon to redshifts $z \lesssim 1.5$. Consequently, we cross-validated the population distribution from \citet{2013ApJ...779...72D} (assuming low metallicity)---as utilized in previous studies such as \citet{10.1093/mnras/stab3298}---against the observational constraints of \citet{theligoscientificcollaboration2025gwtc40populationpropertiesmerging}.

\begin{figure}[t!]
    \centering
    \includegraphics[width=1\linewidth]{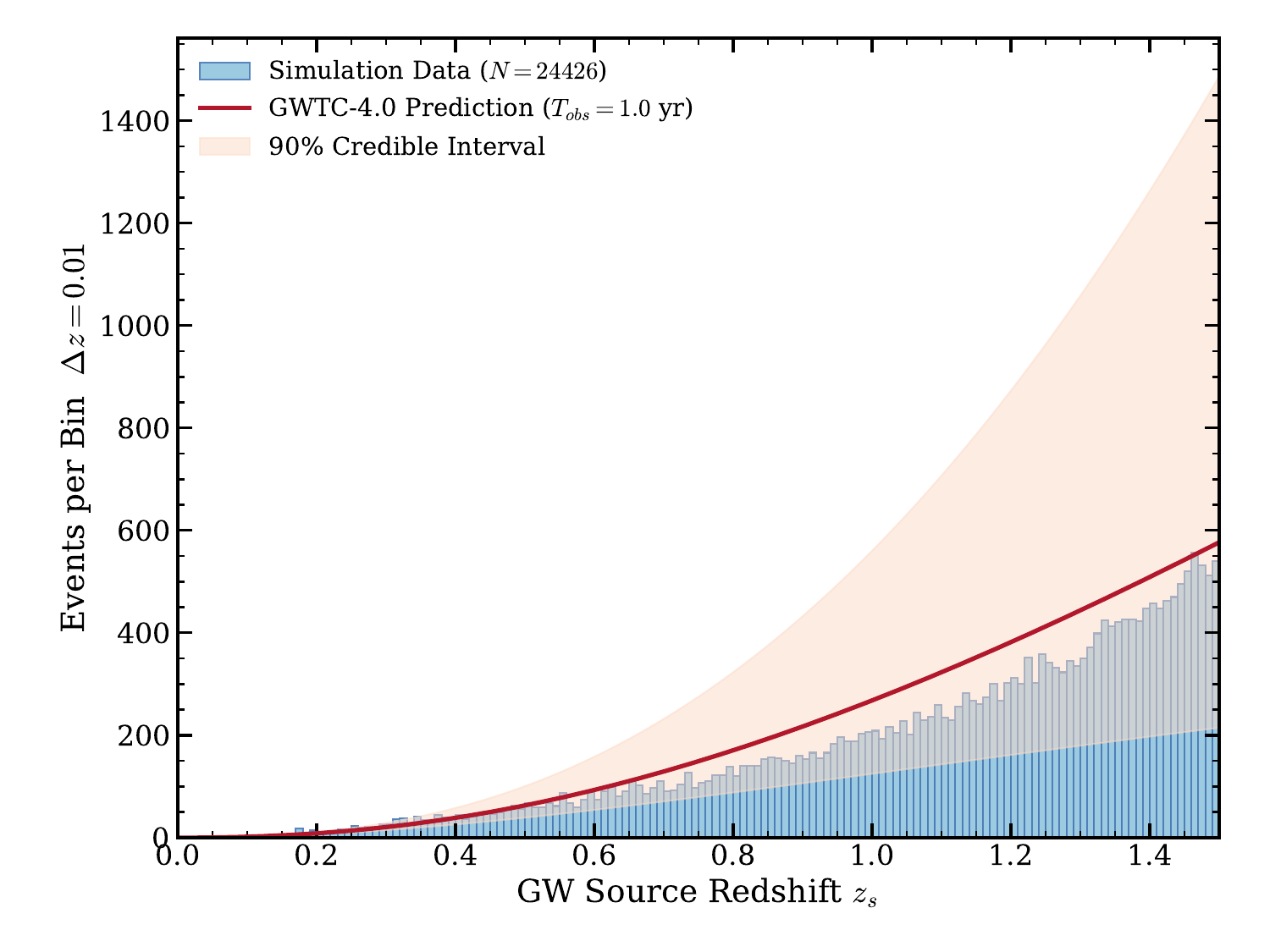}
    \caption{Cross-validation of the redshift distribution from \citet{2013ApJ...779...72D} against the power-law evolution model $\mathcal{R}(z) \propto (1+z)^\kappa$ derived from the \citet{theligoscientificcollaboration2025gwtc40populationpropertiesmerging} population analysis. The horizontal axis represents the GW source redshift, while the vertical axis denotes the number of GW events per unit redshift interval.}
    \label{fig:z_distribution_D2013_GWTC4}
\end{figure}

As shown in \autoref{fig:z_distribution_D2013_GWTC4}, the two distributions are consistent to a reasonable degree within current statistical uncertainties. Therefore, to maintain continuity with earlier research and ensure the robustness of our simulations, we retain the redshift distribution described in \citet{2013ApJ...779...72D}. The merger rates $\dot{n}_0(z_s)$ for all CBC categories (BBH, NSBH, and BNS) are modeled as functions of cosmological redshift according to the prescriptions in \citet{2013ApJ...779...72D}; these models are based on simulations performed using the StarTrack population synthesis code \citep{2008ApJS..174..223B}. Furthermore, we adopted cosmological parameters consistent with \citet{Abe_2025} in our calculations.

\begin{figure}[t!]
    \centering
    \includegraphics[width=1\linewidth]{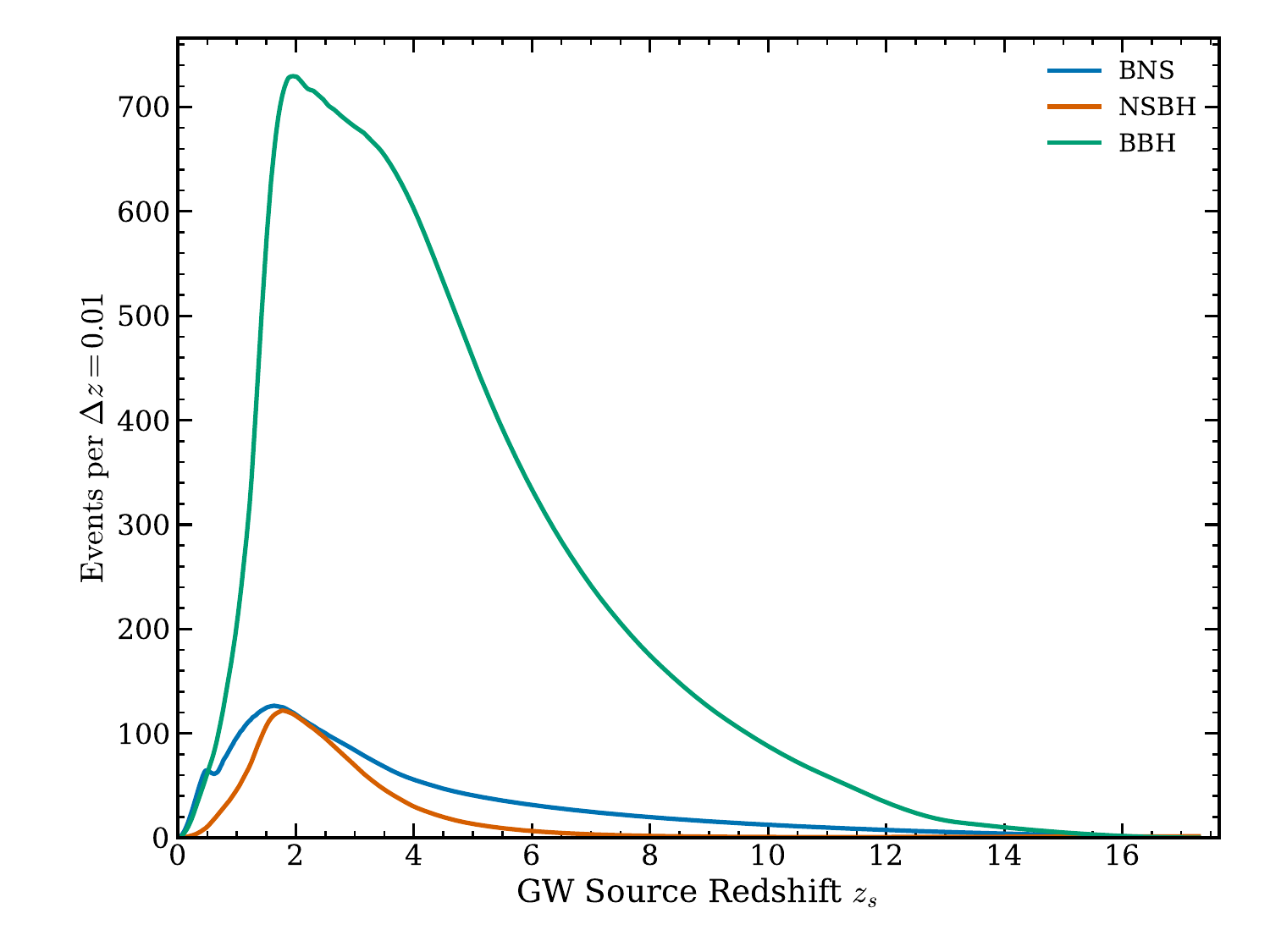}
    \caption{Redshift distributions of GW events for different source types according to Dominik et al. 2013}.

    \label{fig:gw_event}
\end{figure}

To obtain realistic event counts, we performed Poisson sampling based on the theoretically calculated number density of binary merger events per unit redshift. The resulting synthetic dataset comprises approximately $4.0 \times 10^5$ BBH events, $5.7 \times 10^4$ BNS events, and $3.1 \times 10^4$ NSBH events. The theoretical redshift distributions of these populations are presented in \autoref{fig:gw_event}.

\subsection{BBH Population Model}
We construct the synthetic population of BBHs by modeling their redshift evolution, mass spectrum, and spin properties based on the latest observational constraints.

\subsubsection{Primary Mass Distribution}
\citet{10.1093/mnras/stab3298} previously investigated lensed GW event rates assuming two distinct power-law mass distributions: one characterized by a spectral index $\alpha = -2.3$ with component masses $m_{1,2} \in [5, 50]\,M_\odot$, and another with $\alpha = -1$ for $m_{1,2} \in [5, 85]\,M_\odot$. However, these distributions impose sharp cutoffs at the mass boundaries and fail to capture observed substructures, specifically the primary mass peaks located at approximately $10\,M_\odot$ and $35\,M_\odot$. Consequently, we adopt the ``Broken Power Law + 2 Peaks'' model, established as the fiducial model by the LVK Collaboration in \citet{theligoscientificcollaboration2025gwtc40populationpropertiesmerging}, to describe the primary mass distribution of BBHs.

In this framework, the probability density function (with the hyperparameter set $\Lambda$ dictating the shape of the distribution) of the primary mass, $\pi(m_1|\Lambda)$, is modeled as a mixture consisting of a broken power-law backbone and two truncated Gaussian components, incorporating a smoothing function at the low-mass end to model the turn-on behavior. The primary mass distribution is given by:
\begin{equation}
\begin{aligned}
\pi(m_1|\mathbf{\Lambda}) \propto \Bigg[ & \lambda_0 p_{\mathrm{BP}}(m_1 | \alpha_1, \alpha_2, m_{\mathrm{break}}, m_{1,\mathrm{low}}, m_{\mathrm{high}}) \\
& + \lambda_1 \mathcal{N}_{lt}(m_1 | \mu_1, \sigma_1, \mathrm{low} = m_{1,\mathrm{low}}) \\
& + (1 - \lambda_0 - \lambda_1)\mathcal{N}_{lt}(m_1 | \mu_2, \sigma_2, \mathrm{low} = m_{1,\mathrm{low}}) \Bigg] \\
& \times S(m_1 | m_{1,\mathrm{low}}, \delta_{m,1}) ,
\end{aligned}
\end{equation}

where $p_{\mathrm{BP}}(m_1)$ describes the continuous broken power-law spectrum, and $\mathcal{N}_{\mathrm{lt}}$ denotes left-truncated normal distributions representing the two overdensities located at $\sim 10\,M_\odot$ and $\sim 35\,M_\odot$. The parameters $\lambda_0$ and $\lambda_1$ are the mixing fractions determining the relative weights of each component. The term $S(m_1)$ is a smoothing function governing the turn-on behavior at low masses; its explicit functional form is detailed in Appendix B.3 (Binary Black Hole Mass Models) of the  \citet{theligoscientificcollaboration2025gwtc40populationpropertiesmerging}. The hyperparameter set $\mathbf{\Lambda}$ fully characterizes the aforementioned mixture model structure, specifically including: the spectral indices $\alpha_1$ and $\alpha_2$ characterizing the two slopes of the continuous broken power-law backbone, along with its break location $m_{\mathrm{break}}$ and the upper mass truncation $m_{\mathrm{high}}$; the parameters $\mu_1, \sigma_1$ and $\mu_2, \sigma_2$ specifying the locations and widths of the two left-truncated Gaussian peaks; the threshold $m_{1,\mathrm{low}}$ and the taper scale $\delta_{m,1}$ governing the global lower mass truncation and the width of the smoothing transition interval; as well as the mixing fractions $\lambda_0$ and $\lambda_1$ strictly regulating the relative statistical weights of the power-law and the first Gaussian components, respectively (with the second Gaussian component implicitly assigned a weight of $1 - \lambda_0 - \lambda_1$).

\subsubsection{Conditional Mass Ratio Distribution}
Once the primary mass $m_1$ is sampled, the secondary mass $m_2$ is derived from the conditional mass ratio distribution $p(q|m_1)$, where $q = m_2/m_1$. Following the \citet{theligoscientificcollaboration2025gwtc40populationpropertiesmerging} fiducial model, the mass ratio is modeled as a power law bounded by a smoothing function at the low-mass end to ensure physical continuity. The PDF is given by:
\begin{equation}
    p(q|m_1, \beta_q, m_{\mathrm{min}}, \delta_m) \propto q^{\beta_q} S(m_1 q | m_{\mathrm{min}}, \delta_m),
    \label{Conditional Mass Ratio}
\end{equation}
where the spectral index $\beta_q$ governs the preference for symmetric ($\beta_q > 0$) or asymmetric ($\beta_q < 0$) binaries. The term $S(m_1 q | \dots)$ represents the same smoothing function used for the primary mass, here applied to the secondary mass $m_2 = m_1 q$ to model the low-mass turn-on behavior near the minimum mass threshold $m_{\mathrm{min}}$.
\begin{table*}[t!]
    \centering
    \caption{Summary of the intrinsic parameter models adopted for the BBH, BNS, and NSBH populations.}
    \label{tab:intrinsic_params}
    \setlength{\extrarowheight}{3pt}
    \begin{tabular}{llll}
          \hline\hline
        \textbf{Param.} & \textbf{BBH} & \textbf{BNS} & \textbf{NSBH} \\
        \hline
        $m_1$
        & \parbox[t]{4.8cm}{\textbf{Broken Power Law+2 Peaks} \\ $ m_1 \in [3, 300] \, M_\odot$}
        & \parbox[t]{4.8cm}{\textbf{Peak Model} \\ $m_1 \in [1.0, 2.4] \, M_\odot$}
        & \parbox[t]{4.8cm}{\textbf{Truncated Power Law} \\ $ m_1 \in [2, 20] \, M_\odot$} \\
        \hline
        $q$
        & \parbox[t]{4.8cm}{\textbf{Power Law + Smoothing} \\ $q \in [5.0/m_1, 1]$}
        & \parbox[t]{4.8cm}{\textbf{Power Law} \\ $q \in [1.0/m_1, 1]$}
        & \parbox[t]{4.8cm}{\textbf{Truncated Gaussian} \\ $q \in [1.0/m_{\mathrm{BH}}, q_{\mathrm{max}}]$} \\
        \hline
        $\chi$
        & \parbox[t]{4.8cm}{\textbf{Truncated Normal} \\ $\chi \in (0, 1)$}
        & \parbox[t]{4.8cm}{\textbf{Uniform} \\ $\chi \in (0, 0.05)$}
        & \parbox[t]{4.8cm}{\textbf{BH:} Beta distribution \\ \textbf{NS:}  $ U(0, 0.05)$} \\
        \hline
        $\cos \theta$
        & \multicolumn{3}{l}{\parbox[t]{14.4cm}{\centering Aligned with Orbital Angular Momentum ($\cos\theta = 1$)}} \\
        \hline\hline
    \end{tabular}
    \vspace{0.8em}
    \begin{minipage}{\textwidth}
        \footnotesize
        \textbf{Note:}
        The BBH population adopts the fiducial \textit{Broken Power Law + 2 Peaks} mass model (Appendix B.5.1) and the \textit{Gaussian Component Spins} model (Appendix B.3) from \citet{theligoscientificcollaboration2025gwtc40populationpropertiesmerging}.
        For BNS events, the primary mass follows the \textit{Peak model} (Appendix B.2), with a low-spin assumption ($\chi \le 0.05$) \citep{theligoscientificcollaboration2025gwtc40populationpropertiesmerging}.
        The NSBH population relies on the hierarchical framework of \citet{10.1093/mnras/stac3052}, modeling black hole spins with a Beta distribution and mass ratios with a truncated Gaussian.
        We assume spins are perfectly aligned with the orbital angular momentum for all sources.
    \end{minipage}
\end{table*}

\begin{table}[htbp]
    \centering
    \caption{Summary of the extrinsic parameters common to all GW sources.}
    \label{tab:extrinsic_params}
    \renewcommand{\arraystretch}{1.3}
    \begin{tabular}{ll}
       \hline  \hline
        \textbf{Parameter} & \textbf{Distribution \& Range} \\
        \hline
        Redshift ($z$)
        & \citet{2013ApJ...779...72D} \\
        & Range: $z \in [0, 21.5]$ \\
         \hline
        Right Ascension ($\alpha$)
        & \textbf{Uniform} \\
        & $\alpha \in [0, 2\pi]$ \\
         \hline
        Declination ($\delta$)
        & \textbf{Cosine-Uniform (Isotropic)} \\
        & $\delta \in [-\pi/2, \pi/2]$ \\
         \hline
        Inclination ($\iota$)
        & \textbf{Sine-Uniform (Isotropic)} \\
        & $\iota \in [0, \pi]$ \\
       \hline
        Polarization ($\psi$)
        & \textbf{Uniform} \\
        & $\psi \in [0, \pi]$ \\
         \hline
        Coalescence Phase ($\phi_c$)
        & \textbf{Uniform} \\
        & $\phi_c \in [0, 2\pi]$ \\
    \hline  \hline
    \end{tabular}
    \vspace{0.5em}
    \begin{minipage}{\linewidth} 
        \footnotesize
        \textbf{Note:}
        The redshift distribution follows the population synthesis model of \citet{2013ApJ...779...72D}, covering a broad redshift range up to $z \sim 21.5$.
        The sky location parameters (Right Ascension $\alpha$, Declination $\delta$) and binary orientation parameters (Inclination $\iota$, Polarization $\psi$, Phase $\phi_c$) are modeled assuming complete isotropy in space, with no preferred directions.
    \end{minipage}
\end{table}

\subsubsection{Spin Distribution}
We assume that the dimensionless spin magnitudes, $\chi_{1,2}$, of the primary and secondary black holes are independent and identically distributed (IID). Following the ``Gaussian Component Spins'' model outlined in Appendix B.5.1 of \citet{theligoscientificcollaboration2025gwtc40populationpropertiesmerging}, the spin magnitudes follow a truncated normal distribution within the interval $[0, 1]$. The PDF is given by:
\begin{equation}
    \pi(\chi | \mu_\chi, \sigma_\chi) = \mathcal{N}_{[0,1]}(\chi | \mu_\chi, \sigma_\chi),
\end{equation}
where $\mathcal{N}_{[0,1]}$ denotes a normal distribution truncated to the range $[0,1]$, characterized by the mean $\mu_\chi$ and standard deviation $\sigma_\chi$.

To simplify the analysis and focus on isolated binary evolution where precession effects are negligible, we assume the black hole spins are perfectly aligned with the orbital angular momentum (i.e., $\chi_{1,x} = \chi_{1,y} = \chi_{2,x} = \chi_{2,y} = 0$). Consequently, the spin tilt angles are fixed at $\theta_{1,2} = 0$, implying $\cos\theta = 1$.

\subsection{BNS Population Model}
For BNS merger events, we adopted the Peak model for the primary mass distribution, as outlined in Appendix B.2 of the \citet{theligoscientificcollaboration2025gwtc40populationpropertiesmerging} catalog. 
The Peak model introduces a Gaussian component to describe the mass distribution:
\begin{equation}
    \pi(m|\Lambda) \propto
    \begin{cases}
    \exp\left[-\frac{(m-\mu)^2}{2\sigma^2}\right] & \text{if } m_{\mathrm{min}} \le m \le m_{\mathrm{max}}, \\
    0 & \text{otherwise}.
    \end{cases}
\end{equation}
For the conditional mass ratio distribution, we employed a power-law model, where the mass ratio is defined as $q = m_2/m_1$:
\begin{equation}
    p(q|\beta, m_1, m_{\mathrm{min}}) \propto
    \begin{cases}
    q^\beta & \text{if } q_{\mathrm{min}}(m_1) \le q \le 1, \\
    0 & \text{otherwise}.
    \end{cases}
\end{equation}
Here, the distribution reduces to a uniform distribution when the spectral index $\beta = 0$. The lower limit of the mass ratio is determined by the minimum component mass threshold ($m_2 \ge m_{\mathrm{min}}$), such that $q_{\mathrm{min}} = m_{\mathrm{min}}/m_1$. The secondary mass is consequently given by $m_2 = q m_1$.

Given that BNS systems are typically expected to exhibit low spins, we adopt a low-spin assumption. Specifically, we model the dimensionless spin magnitudes, $\chi_1$ and $\chi_2$, using a uniform distribution within the interval $[0, 0.05]$. Consistent with our assumption of negligible precession, we assume the spins are aligned with the orbital angular momentum.

\subsection{NSBH Population Model}
Given that \citet{theligoscientificcollaboration2025gwtc40populationpropertiesmerging} has not yielded definitive new NSBH events, we adopt the hierarchical model framework established by \citet{10.1093/mnras/stac3052} based on the GWTC-3 catalog. The mass model for the NSBH population assumes an independent distribution for the black hole mass, $m_{\mathrm{BH}}$, and a conditional distribution for the mass ratio, $q$. The black hole mass $m_{\mathrm{BH}}$ follows a truncated power-law distribution:
\begin{equation}
    \pi(m_{\mathrm{BH}} | \alpha, m_{\mathrm{BH,min}}, m_{\mathrm{BH,max}}) \propto m_{\mathrm{BH}}^{-\alpha}.
\end{equation}
In this analysis, the upper limit for the black hole mass is set to $20\,M_\odot$ to explicitly exclude events with massive secondary companions, such as GW190814.

The mass ratio, defined as $q = m_{\mathrm{NS}} / m_{\mathrm{BH}}$, is modeled by a truncated Gaussian distribution:
\begin{equation}
    p(q | \mu, \sigma, q_{\mathrm{min}}, q_{\mathrm{max}}) \propto \frac{1}{\sigma\sqrt{2\pi}} \exp\left( -\frac{(q - \mu)^2}{2\sigma^2} \right).
\end{equation}
Since $q$ is a conditional distribution dependent on $m_{\mathrm{BH}}$, the boundaries $[q_{\mathrm{min}}, q_{\mathrm{max}}]$ are calculated dynamically to ensure the neutron star mass remains within a physically reasonable range. Specifically, the lower and upper bounds are defined as:
\begin{align}
    q_{\mathrm{min}} &= 1\,M_\odot / m_{\mathrm{BH}}, \\
    q_{\mathrm{max}} &= \min(m_{\mathrm{NS, max}} / m_{\mathrm{BH}}, 1).
\end{align}
Here, the minimum neutron star mass is fixed at $1\,M_\odot$, while the maximum mass, $m_{\mathrm{NS, max}}$, is treated as a free parameter. Consequently, the neutron star mass is derived via $m_{\mathrm{NS}} = q \times m_{\mathrm{BH}}$.

The dimensionless spin magnitude of the black hole is assumed to follow a Beta distribution:
\begin{equation}
    \pi(\chi_{\mathrm{BH}} | \alpha_{\chi}, \beta_{\chi}) \propto \frac{\chi_{\mathrm{BH}}^{\alpha_{\chi}-1}(1-\chi_{\mathrm{BH}})^{\beta_{\chi}-1}}{B(\alpha_{\chi}, \beta_{\chi})},
\end{equation}
where $\alpha_\chi$ and $\beta_\chi$ are hyperparameters determining the shape of the distribution; specifically, $\alpha_\chi < \beta_\chi$ implies a preference for low spins. This distribution is defined over the interval $[0, 1]$. For the neutron star spin, consistent with our BNS model, we adopt a uniform distribution for the spin magnitude in the range $[0, 0.05]$. To neglect precession effects, we assume all spins are aligned with the orbital angular momentum.

\section{Deflector Model}

Previous simulations of lensed GW events have largely relied on simplified lens models, such as the Singular Isothermal Sphere (SIS) and Singular Isothermal Ellipsoid (SIE). Since these models are restricted to galactic scales, we aim to incorporate the gravitational influence of galaxy groups and clusters to generate a more comprehensive population of lensed GW signals.

In previous works, such as \citet{10.1093/mnras/stab3298}, the lensing probability was determined by calculating the optical depth at various redshifts based on the SIS model. A probabilistic approach was adopted wherein a signal was deemed ``lensed'' if its optical depth exceeded a random number drawn from a uniform distribution between 0 and 1. To establish a more realistic criterion for identifying lensed events, we employ the deflector model presented by \citet{Abe_2025}, which is outlined below.

\citet{Abe_2025} introduced a comprehensive lens system comprising five distinct components: the host dark matter halo, the central galaxy residing at its core, the population of dark matter subhalos surrounding the host, the satellite galaxies hosted within these subhalos, and the external shear induced by mass distributions outside the primary lens system. By constructing this composite mass model, we can achieve a seamless transition across mass scales---from single galaxies to galaxy groups and clusters---thus allowing for a complete assessment of the impact of massive perturbers on GW events.

\subsection{Host dark matter halos and Central galaxies}

First, \citet{Abe_2025} employ the halo mass function proposed by \citet{Tinker_2008} to derive the population distribution of host dark matter halos. They adopt the virial mass definition for the halo mass. The mass-density distribution of the host halo is described by the NFW profile \citep{1996ApJ...462..563N}:
\begin{equation}
    \rho_{\mathrm{hh}}(r) = \frac{\rho_{\mathrm{s,hh}}}{(r/r_{\mathrm{s,hh}})(1+r/r_{\mathrm{s,hh}})^2},
\end{equation}
where $\rho_{\mathrm{s,hh}}$ is the characteristic density and $r_{\mathrm{s,hh}}$ is the scale radius. For the concentration parameter, defined as $c_{\mathrm{hh}} = r_{\mathrm{vir,hh}} / r_{\mathrm{s,hh}}$, they adopt the mass--concentration relation presented in \citet{Diemer_2015} and updated by \citet{Diemer_2019}, assuming that the concentration parameter follows a log-normal distribution. To incorporate the effect of ellipticity $e_{\mathrm{hh}}$, the radial coordinate is transformed into an elliptical coordinate system. The ellipticity $e_{\mathrm{hh}}$ is assumed to follow a truncated normal distribution, with the mean ellipticity varying as a function of halo mass according to the relation provided by \citet{10.1093/mnras/staa1479}. The position angle is drawn from a uniform distribution between $-180^\circ$ and $180^\circ$.

For the central galaxy located within the host halo, the modeling approach is as follows: The stellar mass is derived using the fitting formula for the mean stellar mass fraction, $M_{\mathrm{c}*} = f_{\mathrm{c}*} M_{\mathrm{hh}}$, provided in Appendix J of \citet{10.1093/mnras/stz1182}. A scatter following a log-normal distribution is assumed for the stellar mass--halo mass relation. The stellar component is then described using an elliptical Hernquist profile \citep{1990ApJ...356..359H}:
\begin{equation}
    \rho_{\mathrm{c}*}(\nu_{\mathrm{c}*}) = \frac{f_{\mathrm{c}*}M_{\mathrm{hh}}}{2\pi(\nu_{\mathrm{c}*}/r_{\mathrm{b,cen}})(\nu_{\mathrm{c}*} + r_{\mathrm{b,cen}})^3},
\end{equation}
where $\nu_{\mathrm{c}*}$ is the radial distance projected along the line of sight $r_{\mathrm{b,cen}}$ represents the scale
radius of the galaxy, related to the effective radius by $r_{\mathrm{b,cen}} = 0.551 r_{\mathrm{e,cen}}$. 
The effective radius $r_{\mathrm{e,cen}}$ is determined using the size--mass relation for quiescent galaxies derived by \citet{vdW2024} based on JWST observations, assuming the scatter of $r_{\mathrm{e,cen}}$ follows a log-normal distribution. Regarding the shape of the central galaxy, the ellipticity $e_{\mathrm{c}*}$ is modeled using a truncated normal distribution with a fixed mean of 0.3 and a dispersion of 0.16, consistent with SDSS observations. Furthermore, to account for the alignment between the galaxy and its halo, the position angle of the central galaxy is modeled as a Gaussian distribution centered on the position angle of its host halo ($\phi_{\mathrm{hh}}$).

\subsection{Subhalos and External shear}

In the \texttt{SL-Hammocks} framework, the subhalo population is modeled using the analytic approach proposed by \citet{Oguri_2020}. This model is grounded in the extended Press--Schechter theory \citep{1991ApJ...379..440B,10.1093/mnras/248.2.332,1993MNRAS.262..627L} but incorporates the additional effects of tidal stripping and dynamical friction. The subhalo mass function at a given redshift $z$ is expressed as:
\begin{equation}
    \frac{\mathrm{d}N_{\mathrm{sh}}}{\mathrm{d}M_{\mathrm{sh}}} = f_{\mathrm{df}} \frac{\mathrm{d}N_{\mathrm{sh}}}{\mathrm{d}M_{\mathrm{f}}} \frac{\mathrm{d}M_{\mathrm{f}}}{\mathrm{d}M_{\mathrm{sh}}}.
\end{equation}
This formulation derives the current subhalo mass function by transforming the progenitor abundance at the epoch of accretion. It includes a survival fraction, $f_{\mathrm{df}}$, which accounts for dynamical friction, and a 
term, $\mathrm{d}M_{\mathrm{f}}/\mathrm{d}M_{\mathrm{sh}}$, which maps the mass at accretion ($M_{\mathrm{f}}$) to the current mass ($M_{\mathrm{sh}}$) following tidal stripping.

Regarding the spatial distribution, \citet{Abe_2025} assume that subhalos trace the mass distribution of the host halo (modeled with an NFW profile). Subhalos are populated within the host halo using inverse transform sampling based on the radial cumulative distribution function, $P_{\mathrm{sh}}(x_{\mathrm{hh}}, M_{\mathrm{f}})$. $x_{\mathrm{hh}} = r/r_{vir}$ is dimensionless radial distance---see \citet{Abe_2025} for the notation we adopted.

Although subhalos are theoretically expected to exhibit truncated NFW profiles due to tidal stripping, the model adopts a non-truncated elliptical NFW profile for the actual lensing calculations to ensure numerical efficiency. Crucially, the lensing properties are computed using the subhalo mass at the accretion epoch ($M_{\mathrm{f}}$) rather than the stripped mass ($M_{\mathrm{sh}}$). This approximation is justified because strong lensing effects are dominated by the central mass distribution (near the Einstein radius), which remains largely robust against the tidal stripping that affects the outer halo.

The concentration parameter for subhalos, $\bar{c}_{\mathrm{sh}}$, is determined using the fitting formula from \citet{10.1093/mnras/staa069}, which is specialized for subhalo structures. Since this relation is valid only for $M_{\mathrm{sh}} < 10^{12} M_{\odot}$, the model adopts the larger value between the \citet{10.1093/mnras/staa069} result and the field--halo relation from \citet{Diemer_2019} (DKJ) for higher masses.

Furthermore, a satellite galaxy is assigned to the center of each subhalo. Its properties are derived using the stellar mass--halo mass relation from \citet{10.1093/mnras/stz1182}, adopting an elliptical Hernquist profile consistent with the treatment of central galaxies. Consequently, when computing the lensing effect of a specific subhalo, the model incorporates contributions from the host halo, the central galaxy, the subhalo itself, the satellite galaxy, and external shear. Further details on the model implementation can be found in \citet{Abe_2025}. \texttt{SL-Hammocks} adopts the truncated normal distribution form from OM10+ for external shear but improves the treatment of the dispersion, $\sigma_{\gamma_{\mathrm{ext}}}$, by introducing redshift dependence. For lensing calculations, the multi-body lensing problem is decomposed into independent computational units dominated by host halos and subhalos, respectively. This approach significantly optimizes the computational efficiency for generating large-scale mock catalogs while maintaining physical accuracy.

\section{Methods}

\subsection{Synthetic Population Generation}

To generate a synthetic BBH population consistent with observational constraints and to fully quantify the statistical uncertainties arising from the finite number of observations, we employed a Monte Carlo sampling method based on the Posterior Predictive Distribution (PPD). The specific procedure is as follows:

\begin{enumerate}
    \item \textbf{Hyperparameter Posterior Extraction:}
    Our simulation relies not on a single ``best-fit'' parameter set, but rather utilizes the full chain of hyperparameter posterior samples provided by the LIGO-Virgo-KAGRA (LVK) Collaboration for \citet{Abe_2025}, specifically the publicly available data release on Zenodo \citep{LVK_Zenodo_GWTC4}. For each simulated GW event (or batch of events), we dynamically draw a corresponding vector of population hyperparameters, $\Lambda$ (including mass model parameters such as $\alpha, m_{\mathrm{break}}, \mu_{\mathrm{peak}}$ and spin model parameters $\mu_\chi, \sigma_t$). This step ensures that our synthetic population naturally incorporates the confidence intervals of the population inference derived from current observational data.

    \item \textbf{Mass Sampling (Rejection Sampling and Inverse Transform Methods):}
    For the primary mass $m_1$, the  PDF of the ``Broken Power Law + 2 Peaks'' model is analytically complex. Consequently, we implemented a Rejection Sampling algorithm. We first generated candidate samples from an envelope distribution covering the range $[m_{\mathrm{min}}, m_{\mathrm{max}}]$. These candidates were then accepted or rejected based on an acceptance probability derived from the target mass distribution function, effectively filtering the samples to match the fiducial model.

    The generation of the secondary mass $m_2$ depends on the conditional mass ratio distribution $p(q|m_1)$. According to the model described in Equation~\ref{Conditional Mass Ratio}, the mass ratio $q$ follows a power-law distribution modulated by a smoothing function. In our numerical implementation, we adopted a two-step strategy combining \textit{Inverse Transform Sampling} and \textit{Rejection Sampling}:

    \begin{itemize}
        \item \textit{Step 1 (Power-Law Component):} Based on the extracted spectral index $\beta_q$, we used the Inverse Transform Sampling method to draw a candidate mass ratio $q_{\mathrm{try}}$ from the pure power-law distribution $p(q) \propto q^{\beta_q}$. The tentative secondary mass was then calculated as $m_{2,\mathrm{try}} = m_1 \cdot q_{\mathrm{try}}$.

        \item \textit{Step 2 (Smoothed Truncation):} To enforce the physical constraints at the low-mass end, we applied the smoothing function $S(m)$ as an acceptance probability factor. We generated a uniform random number $u \sim U[0,1]$ and compared it with $S(m_{2,\mathrm{try}})$. If $u < S(m_{2,\mathrm{try}})$, the sample was accepted, determining the final secondary mass $m_2 = m_{2,\mathrm{try}}$; otherwise, the sample was rejected and the process repeated. This procedure accurately simulates the ``turn-on'' behavior of the secondary mass near the minimum mass threshold, ensuring strict adherence to the physical model.
    \end{itemize}

    \item \textbf{Spin Sampling:}
    For the spin parameters, consistent with the isolated binary evolution scenario, we focused solely on the spin magnitudes. Using the extracted hyperparameters $\mu_\chi$ and $\sigma_\chi$, we sampled the dimensionless spin magnitudes $\chi$ from a truncated normal distribution within the interval $[0,1]$. Regarding the spin orientation, we adopted a perfectly aligned assumption by fixing the spin tilt cosines to $\cos\theta = 1$ for all events, thereby neglecting precession effects.
\end{enumerate}

Through this procedure, we generated a synthetic catalog comprising approximately $4.0 \times 10^5$ BBH merger events. This catalog not only adheres to the latest astrophysical population models but also propagates the statistical errors from the model inference by traversing the posterior sample space.

\subsection{GW Equivalent Luminosity Function}

We first aim to construct a joint  PDF of redshift and signal-to-noise ratio (SNR) for GW sources. Based on the binary merger event rate, we generate the number of GW events within discrete redshift intervals. For the GW events within a given redshift bin, their specific redshifts are determined by sampling from a Poisson distribution. Subsequently, we generate a set of source parameters for each GW signal according to their respective relevant distributions. Finally, we calculate the SNR of each signal as observed by the detector network. This procedure yields the PDF of GW sources as a function of redshift and their corresponding SNRs.

Since the simulation software we employ, SL-Hammocks, is developed for optical surveys, it can directly accept the luminosity functions of quasars and supernovae as inputs. In the electromagnetic band, a luminosity function is essentially the PDF of celestial objects over luminosity, typically utilizing absolute magnitude as the input parameter. In SL-Hammocks, absolute magnitude serves merely as a baseline to deduce apparent magnitude, which is then used to filter out background sources that meet the observational limits of optical telescopes (e.g., the apparent magnitude limit of $\tilde{m}=23.3$ set for LSST). However, GW observations directly measure wave amplitude and SNR, parameters that cannot be directly fed into this code. To robustly utilize this open-source software and avoid introducing unknown systematic errors by directly modifying the underlying code, we assign an ``equivalent absolute magnitude'' to the SNR of each GW source, thereby constructing an ``equivalent luminosity function'' for GWs.

Following conventions of SDSS-III,  we first establish the conversion between the apparent magnitude $m$ and the corresponding flux $F$ based on Pogson's Equation:
\begin{equation}
    m = -2.5 \log_{10}(F) + 22.5,
    \label{eq:mag_conversion}
\end{equation}
where $F$ is in units of nanomaggies $nMgy \approx 3.631\,;10^{-6}\;Jy$, and $22.5$ corresponds to the photometric zero point of the system. Given that the limiting apparent magnitude of the optical telescope in SL-Hammocks is set to $\tilde{m}=23.3$, substituting this into Equation~\ref{eq:mag_conversion} yields a corresponding flux threshold of $\tilde{F}=0.478$.

Given that the SNR is inversely proportional to the luminosity distance ($\text{SNR} \propto d_L^{-1}$), while the flux $F$ of an optical image is inversely proportional to the square of the luminosity distance ($F \propto d_L^{-2}$), we can establish the following analogy between the two:
\begin{equation}
    F^* = \tilde{F} \left( \frac{\rho_{\mathrm{intr}}}{\rho_{\mathrm{threshold}}} \right)^2,
    \label{eq:snr_to_flux}
\end{equation}
where $F^*$ represents the equivalent flux corresponding to any given GW signal, $\rho_{\mathrm{intr}}$ denotes the intrinsic SNR, and $\rho_{\mathrm{threshold}}$ represents the SNR threshold we use to select candidate lensed events. Here, we conservatively set $\rho_{\mathrm{threshold}} = 0.1$. The rationale for this choice is multifaceted: we must account for the time delays introduced by gravitational lensing, the variation in the detector's antenna pattern sensitivity to signals arriving at different times, and the magnification bias inherent to lensed GW events. Setting a sufficiently low intrinsic SNR threshold ensures that we do not omit events that are intrinsically very weak but have undergone extreme magnification. By anchoring our conversion with Equation~\ref{eq:snr_to_flux}, the software will automatically filter out extremely weak events that fail to meet our detection requirements even after being magnified by a lens.

Finally, we utilize the standard conversion formula between absolute and apparent magnitudes:
\begin{equation}
    M^* = m^* - 5 \log_{10}(d_L) + 25,
    \label{eq:absolute_magnitude}
\end{equation}
where $d_L$ is the luminosity distance of the GW source in units of Mpc, $m^*$ represents the equivalent apparent magnitude of the GW source, obtained by substituting the equivalent flux $F^*$ calculated from Equation~\ref{eq:snr_to_flux} back into Equation~\ref{eq:mag_conversion}, and $M^*$ is the corresponding equivalent absolute magnitude. By combining Equations~\ref{eq:mag_conversion}, \ref{eq:snr_to_flux}, and \ref{eq:absolute_magnitude}, we successfully map the redshift and SNR of GW sources into an equivalent luminosity space, subsequently deriving its equivalent luminosity function $\Phi(z_s, M^*)$. This serves strictly as an operational interface for utilizing the SL-Hammocks software. The sole purpose of this function is to act as a bridge, allowing the joint distribution of redshift and SNR of GW events to be fed into the SL-Hammocks lensing event sampling pipeline. Furthermore, considering the unprecedented sensitivity of third-generation detectors, the ``equivalent luminosity'' of GW sources is extremely high compared to quasars; as long as the source and the lens satisfy the geometric alignment conditions, the vast majority of their lensed signals will cross the detection threshold.

\section{Mock Catalog for Lensed GWs}
For BBH events, we generated approximately 400,000 GW sources distributed across a sky area of $40,000\,\mathrm{deg}^2$.
Drawing an analogy from optical gravitational lensing, the central images in intrinsic three-image and five-image systems are typically highly demagnified. Furthermore, in the optical regime, these faint signals are often obscured by the luminous emission from the central lens galaxy, rendering them unobservable under current conditions. Consequently, in our analysis, we systematically discarded the image with the minimum magnification (the central image) from all multi-image systems. After applying the selection criteria of our composite lens model over a one-year observational period, we identified over 750 lensed GW systems with a SNR greater than 0.1. This sample comprises approximately 530 doublets (two-image systems), 30 quadruplets (four-image systems), and 140 single, high-magnification ($\mu > 3$) events.

\subsection{PlusNetworks + 3G Detectors}

We conducted simulations of strongly lensed GW events spanning a ten-year observation period, encompassing three distinct source populations---BBH, BNS, and NSBH---across various detector network configurations, including A+ , ET, CE, and the combined ET+CE network. In our simulations, the A+ detector network comprises two $4\,\mathrm{km}$ Advanced LIGO Plus detectors (located at Hanford and Livingston), a $3\,\mathrm{km}$ Advanced Virgo Plus (AdV+) detector, and the KAGRA detector operating at their design sensitivity. The lower frequency cutoff for integration across all detectors in this network is uniformly set to $10\,\mathrm{Hz}$. The ET is modeled in a triangular configuration, consisting of three nested 
interferometers, each with an arm length of $10\,\mathrm{km}$. Its assumed location 
is the current Virgo site in Cascina, Italy, and we employed the ET-D power spectral density (PSD) template, with an observational frequency range spanning from $1\,\mathrm{Hz}$ to $2048\,\mathrm{Hz}$. Furthermore, the CE adopts the classical L-shaped configuration but operates on a vastly expanded scale, featuring an arm length of $40\,\mathrm{km}$. Located at the existing LIGO Hanford site in Washington, USA, our simulation utilizes the CE1 (Stage 1) amplitude spectral density (ASD) data for this detector, with its frequency range configured from $5\,\mathrm{Hz}$ to $2048\,\mathrm{Hz}$ .The PSD/ASD curves for all future detectors (ET and CE) used in this work are sourced from the LIGO Document Control Center \citep{Abbott_2017}. The resulting yearly event rates are summarized in \autoref{tab:lensed_events_final}.
Our classification framework for lensed GW events generally aligns with standard optical lensing morphologies, categorizing events into doublets, quadruplets, and high-magnification events. Furthermore, in the specific context of GW astronomy, we introduce additional categories that differ from traditional optical classifications, such as central-image detections and subhalo-lensed events.

\begin{table*}[htbp]
    \centering
    \caption{Predicted Yearly Rates of Lensed GW Events for Different Detectors and Source Types}
    \label{tab:lensed_events_final}

    \footnotesize
    \renewcommand{\arraystretch}{1.5}
    \setlength{\tabcolsep}{5pt}

    \begin{tabular}{lcccccccccc}
        \toprule
        \toprule

        & \multicolumn{4}{c}{Multiple images ($\mathrm{yr}^{-1}$)}
        & \multicolumn{2}{c}{Central Image cases ($\mathrm{yr}^{-1}$)}
        & \multicolumn{2}{c}{High-mag ($\mathrm{yr}^{-1}$)}
        & \multirow{2}{*}{$N_{\mathrm{trig}}$} \\

        \cmidrule(lr){2-5} \cmidrule(lr){6-7} \cmidrule(lr){8-9}

        Source
        & $N_{\mathrm{multi}}$
        & $N_{\mathrm{double}}$
        & $N_{\mathrm{quad}}$
        & $N_{\mathrm{sub}}$
        & $N^{\mathrm{CI}}_{3}$
        & $N^{\mathrm{CI}}_{5}$
        & $N_{|\mu|>3}$
        & $N_{|\mu|>10}$
        & (Any $>8$) \\
        \midrule

        \multicolumn{10}{l}{\textbf{Detector: ET + CE}} \\
        \midrule
        BBH
        & $431.50^{+6.67}_{-6.57}$      
        & $395.40^{+6.39}_{-6.29}$      
        & $36.10^{+2.00}_{-1.90}$       
        & $107.10^{+3.37}_{-3.27}$      
        & $20.00^{+1.52}_{-1.41}$       
        & $0.40^{+0.32}_{-0.19}$        
        & $359.30^{+6.09}_{-5.99}$      
        & $15.90^{+1.36}_{-1.26}$       
        & $617.20^{+7.96}_{-7.86}$ \\   

        BNS
        & $4.80^{+0.80}_{-0.69}$       
        & $3.60^{+0.71}_{-0.60}$       
        & $1.20^{+0.46}_{-0.34}$       
        & $1.60^{+0.51}_{-0.40}$       
        & $0.70^{+0.38}_{-0.26}$       
        & $0.00^{+0.18}_{-0.00}$       
        & $16.30^{+1.38}_{-1.28}$      
        & $2.00^{+0.55}_{-0.44}$       
        & $20.00^{+1.52}_{-1.41}$ \\   

        NSBH
        & $13.40^{+1.26}_{-1.16}$      
        & $12.50^{+1.22}_{-1.12}$      
        & $0.90^{+0.41}_{-0.29}$       
        & $3.10^{+0.66}_{-0.55}$       
        & $0.40^{+0.32}_{-0.19}$       
        & $0.00^{+0.18}_{-0.00}$       
        & $15.30^{+1.34}_{-1.24}$      
        & $0.40^{+0.32}_{-0.19}$       
        & $29.70^{+1.83}_{-1.72}$ \\   
        \midrule

        \multicolumn{10}{l}{\textbf{Detector: ET}} \\
        \midrule
        BBH
        & $293.90^{+5.52}_{-5.42}$      
        & $263.10^{+5.23}_{-5.13}$      
        & $30.80^{+1.86}_{-1.75}$       
        & $74.00^{+2.82}_{-2.72}$       
        & $10.50^{+1.13}_{-1.02}$       
        & $0.80^{+0.39}_{-0.28}$        
        & $327.50^{+5.82}_{-5.72}$      
        & $16.90^{+1.40}_{-1.30}$       
        & $556.80^{+7.56}_{-7.46}$ \\   

       BNS
        & $1.60^{+0.51}_{-0.40}$       
        & $1.30^{+0.47}_{-0.36}$       
        & $0.30^{+0.29}_{-0.16}$       
        & $0.50^{+0.34}_{-0.22}$       
        & $0.00^{+0.18}_{-0.00}$       
        & $0.00^{+0.18}_{-0.00}$       
        & $7.20^{+0.95}_{-0.85}$       
        & $1.30^{+0.47}_{-0.36}$       
        & $8.10^{+1.00}_{-0.90}$ \\    

       NSBH
        & $4.20^{+0.75}_{-0.65}$       
        & $3.60^{+0.71}_{-0.60}$       
        & $0.60^{+0.36}_{-0.24}$       
        & $1.20^{+0.46}_{-0.34}$       
        & $0.10^{+0.23}_{-0.08}$       
        & $0.10^{+0.23}_{-0.08}$       
        & $10.60^{+1.13}_{-1.03}$      
        & $0.90^{+0.41}_{-0.29}$       
        & $16.50^{+1.39}_{-1.28}$ \\   
        \midrule

        \multicolumn{10}{l}{\textbf{Detector: CE}} \\
        \midrule
       BBH
        & $294.40^{+5.53}_{-5.43}$      
        & $260.80^{+5.21}_{-5.11}$      
        & $33.60^{+1.93}_{-1.83}$       
        & $75.20^{+2.84}_{-2.74}$       
        & $11.70^{+1.18}_{-1.08}$       
        & $0.60^{+0.36}_{-0.24}$        
        & $328.70^{+5.83}_{-5.73}$      
        & $17.10^{+1.41}_{-1.31}$       
        & $546.80^{+7.50}_{-7.39}$ \\   

        BNS
        & $2.20^{+0.58}_{-0.47}$       
        & $1.80^{+0.53}_{-0.42}$       
        & $0.40^{+0.32}_{-0.19}$       
        & $0.50^{+0.34}_{-0.22}$       
        & $0.00^{+0.18}_{-0.00}$       
        & $0.00^{+0.18}_{-0.00}$       
        & $9.70^{+1.09}_{-0.98}$       
        & $1.10^{+0.44}_{-0.33}$       
        & $11.10^{+1.16}_{-1.05}$ \\   

        NSBH
        & $5.60^{+0.85}_{-0.75}$       
        & $5.10^{+0.82}_{-0.71}$       
        & $0.50^{+0.34}_{-0.22}$       
        & $1.40^{+0.48}_{-0.37}$       
        & $0.00^{+0.18}_{-0.00}$       
        & $0.10^{+0.23}_{-0.08}$       
        & $10.90^{+1.15}_{-1.04}$      
        & $0.50^{+0.34}_{-0.22}$       
        & $17.90^{+1.44}_{-1.34}$ \\   
        \midrule

        \multicolumn{10}{l}{\textbf{Detector: PlusNetworks (A+)}} \\
        \midrule
        BBH
        & $2.80^{+0.64}_{-0.53}$       
        & $1.70^{+0.52}_{-0.41}$       
        & $1.10^{+0.44}_{-0.33}$       
        & $0.90^{+0.41}_{-0.29}$       
        & $0.00^{+0.18}_{-0.00}$       
        & $0.00^{+0.18}_{-0.00}$       
        & $15.90^{+1.36}_{-1.26}$      
        & $5.40^{+0.84}_{-0.73}$       
        & $17.30^{+1.42}_{-1.31}$ \\   

        BNS
        & $0.00^{+1.84}_{-0.00}$
        & $0.00^{+1.84}_{-0.00}$
        & $0.00^{+1.84}_{-0.00}$
        & $0.00^{+1.84}_{-0.00}$
        & $0.00^{+1.84}_{-0.00}$
        & $0.00^{+1.84}_{-0.00}$
        & $0.00^{+1.84}_{-0.00}$
        & $0.00^{+1.84}_{-0.00}$
        & $0.00^{+1.84}_{-0.00}$ \\

        NSBH
        & $0.00^{+1.84}_{-0.00}$
        & $0.00^{+1.84}_{-0.00}$
        & $0.00^{+1.84}_{-0.00}$
        & $0.00^{+1.84}_{-0.00}$
        & $0.00^{+1.84}_{-0.00}$
        & $0.00^{+1.84}_{-0.00}$
        & $0.00^{+1.84}_{-0.00}$
        & $0.00^{+1.84}_{-0.00}$
        & $0.00^{+1.84}_{-0.00}$ \\

        \bottomrule
        \bottomrule
    \end{tabular}

    \vspace{0.5em}
    \begin{minipage}{\linewidth}
        \footnotesize
        \textbf{Note:}
        The predictions for lensed GW event rates presented here assume the Planck 2018 cosmological parameters and a standard isolated binary evolution model under low-metallicity conditions.
        The ``Multiple images'' category represents strongly lensed systems ($N_{\mathrm{multi}} = N_{\mathrm{double}} + N_{\mathrm{quad}}$), where $N_{\mathrm{double}}$ denotes systems with two detectable signal copies from the same source, and $N_{\mathrm{quad}}$ refers to systems with at least three detectable signal copies. This category also includes subhalo-dominated lensing systems, denoted as $N_{\mathrm{sub}}$.
        The ``Central Image'' category ($N^{\mathrm{CI}}$) identifies special systems where the typically demagnified central image becomes observable; specifically, $N_3^{\mathrm{CI}}$ and $N_5^{\mathrm{CI}}$ correspond to systems where a total of three and five GW signal copies are detected, respectively.
        ``High-mag'' refers to high-magnification systems (including both single and multiple images) where the maximum magnification exceeds 3 ($|\mu|_{\max} > 3$), regardless of whether all individual images are detected.
        $N_{\mathrm{trig}}$ represents the set of lensed events where at least one image achieves a SNR greater than 8.
        Uncertainties correspond to the 68\% confidence level (CL). We report the exact asymmetric Poisson confidence intervals \citep{Garwood1936} for all event counts, irrespective of sample size.
    \end{minipage}
     \label{tab:lensed_events_final} 
\end{table*}

First, we begin by constructing an initial candidate population of lensed GW events, pre-selecting those with a SNR greater than 0.1. Subsequently, we account for the effects of \textbf{magnification bias} on signal detectability. For each image in a lensed system, the observed SNR, $\rho_{\mathrm{obs}}$, is derived from the intrinsic SNR, $\rho_{\mathrm{intr}}$, and the magnification factor, $\mu$, via the relation:
\begin{equation}
\rho_{\mathrm{obs}} = \rho_{\mathrm{ori}} \sqrt{|\mu|}.
\label{eq:snr_obs}
\end{equation}
Furthermore, since multiple images from the same source traverse different optical paths, we must incorporate the impact of \textbf{time delays}. During the simulation, each GW source is assigned a randomized base arrival time at Earth, denoted as $t_{\mathrm{base}}$. For a given lensed system, we set the relative time delay of the first-arriving image to zero. The arrival times for subsequent images are then calculated by sorting them according to their relative time delays, such that:
\begin{equation}
t_{\mathrm{arrive}} = t_{\mathrm{base}} + t_{\mathrm{delay}}.
\label{eq:time_arrival}
\end{equation}
This adjustment ensures that the time-dependent detector response is correctly applied to each image. Following these procedures, the candidate population is ready for the final selection of detectable lensed GW events. The specific selection logic for categorizing these events is defined as follows:

\begin{enumerate}
    \item \textbf{Total Triggers ($N_{\mathrm{trig}}$):}
    A lensed GW system is counted towards the total triggered events, $N_{\mathrm{trig}}$, if \textit{any} individual image within the system possesses a SNR greater than the detection threshold (i.e., $\rho > 8$).

    \item \textbf{Double-Image Events ($N_{\mathrm{double}}$):}
    For systems that correspond to an intrinsic three-image configuration, we classify the event as a detectable double if the first two arriving images both satisfy $\rho > 8$.

    \item \textbf{Quadruple-Image Events ($N_{\mathrm{quad}}$):}
    For systems that correspond to an intrinsic five-image configuration, the event is classified as a detectable quad ($N_{\mathrm{quad}}$) if the first three arriving images all satisfy $\rho > 8$.

    \item \textbf{High-Magnification Events:}
    A system is recorded as a high-magnification event if the specific image associated with the maximum magnification factor has $\rho > 8$, and the magnification satisfies $|\mu| > 3$ (or $|\mu| > 10$).

    \item \textbf{Detectable Central Images ($N^{\mathrm{CI}}$):}
    We identify a unique class of events where the central image, typically demagnified and unobservable in the electromagnetic regime, becomes detectable due to the sensitivity of 3G detectors. Specifically, we flag cases where all three images of a doublet system exceed the threshold ($N^{\mathrm{CI}}_{3}$) and rare cases where all five images of a quadruplet system are detectable ($N^{\mathrm{CI}}_{5}$).

    \item \textbf{Subhalo Lensing Events ($N_{\mathrm{sub}}$):}
    Within the multi-image population, we isolate events dominated by dark matter substructures. In the \texttt{SL-Hammocks} pipeline, a system is classified as a subhalo lensing event ($N_{\mathrm{sub}}$) if the subhalo's retained mass fraction following tidal truncation falls within the interval of $(0, 1]$, indicating that the subhalo plays a significant role in the lensing potential.
\end{enumerate}

The statistical results presented in \autoref{tab:lensed_events_final} underscore the profound leap in sensitivity offered by third-generation (3G) detectors, which corresponds to a substantial increase in the detection rate of lensed GW events. While the upcoming A+ network is projected to detect only a marginal number of lensed BBH events ($\sim 2.8\,\mathrm{yr}^{-1}$), the deployment and operation of ET and CE will dramatically elevate this figure. Based on current results, BBH events overwhelmingly dominate the lensed catalog, constituting approximately 95\% of the total detected events within the ET+CE configuration. In contrast, lensed BNS and NSBH events remain comparatively rare, with projected detection rates of $\sim 4.8\,\mathrm{yr}^{-1}$ and $\sim 13.4\,\mathrm{yr}^{-1}$, respectively.

Overall, double-image systems account for 91.5\% of all multiple-image events. Conversely, quadruple-image systems are scarcer, and the complete detection of all four images proves difficult, posing new challenges for future identification algorithms. {Within detectable multi-image systems, the most prevalent configuration is the three-image detection of a quadruplet (an intrinsic five-image system). To ensure a comprehensive assessment, we specifically quantified cases within intrinsic five-image systems where only two images pass the detection threshold. Our 10-year simulation identified only ten such events, indicating that the impact of extreme demagnification on the overall predicted detection rates is negligible.Given that the highly magnified image pair formed near the caustics typically corresponds to the second and third arriving signals—while the first arrival is generally fainter—the rarity of cases where the leading signal remains undetected implies a significant conclusion: if a pair of highly similar signals with a very short time delay (suspected 2nd and 3rd images) is identified in real-time, there is a very high probability that the preceding leading image already exists within the observational data. This finding establishes a crucial observational strategy for conducting targeted retrospective searches to confirm and recover the complete lensing topology.}

Furthermore, lensed events significantly lensed by subhalos exhibit notable detection rates; specifically, the ET+CE network is predicted to detect approximately 110 subhalo-dominated multi-image events annually. These events carry the distinct imprint of small-scale dark matter substructures. Historically, such perturbations have been difficult to identify in electromagnetic observations due to the finite size of background sources. However, the point-like nature of GW sources renders them highly sensitive probes for detecting these perturbations. The accumulation of such a statistical sample would provide unprecedented constraints on the abundance of low-mass dark matter halos and offer a stringent test of the Cold Dark Matter (CDM) paradigm.

Moreover, the detection of central images ($N^{\mathrm{CI}}$), although rare ($\sim 20\,\mathrm{yr}^{-1}$ for $N^{\mathrm{CI}}_3$), represents a unique scientific opportunity. The predicted yield of such events in the 3G era will allow us to directly probe the mass distribution in the innermost regions of galaxies, potentially distinguishing between cuspy (NFW-like) and cored density profiles.

 {It is worth noting that the predicted event rates presented in \autoref{tab:lensed_events_final} are necessarily optimistic. While our simulation  of lensed signals based on strict SNR thresholds, the successful detection of a signal does not guarantee its identification as a lensed event in practice. The authentication pipeline is frequently challenged by various observational and analytical uncertainties. Key limiting factors include parameter degeneracies (e.g., the well-known magnification-distance degeneracy), waveform systematics, complex non-Gaussian detector noise (glitches), and the false-alarm background arising from the chance coincidence of independent, unlensed mergers. Furthermore, unmodeled intrinsic source properties, such as orbital eccentricity or spin precession, can obscure or mimic lensing signatures. Consequently, the actual identification efficiency will inevitably be reduced compared to our theoretical detection rates.}

\subsection{Model Dependencies}

\begin{table*}[t]
    \centering
    \caption{Impact of Isolated Binary Evolution Models on Lensed GW Event Rates}
    \label{tab:lensed_events_models_full_comparison}

    \scriptsize
    \renewcommand{\arraystretch}{1.5}
    \setlength{\tabcolsep}{0pt}

    \begin{tabular*}{\textwidth}{@{\extracolsep{\fill}}ccccccccccc}
        \toprule
        \toprule

        & & \multicolumn{4}{c}{Multiple images ($\mathrm{yr}^{-1}$)}
        & \multicolumn{2}{c}{Central Image cases ($\mathrm{yr}^{-1}$)}
        & \multicolumn{2}{c}{High-mag ($\mathrm{yr}^{-1}$)}
        & \multirow{2}{*}{$N_{\mathrm{trig}}$} \\

        \cmidrule(lr){3-6} \cmidrule(lr){7-8} \cmidrule(lr){9-10}

        Met. & Source
        & $N_{\mathrm{multi}}$
        & $N_{\mathrm{double}}$
        & $N_{\mathrm{quad}}$
        & $N_{\mathrm{sub}}$
        & $N^{\mathrm{CI}}_{3}$
        & $N^{\mathrm{CI}}_{5}$
        & $N_{|\mu|>3}$
        & $N_{|\mu|>10}$
        & (Any $>8$) \\
        \midrule

        \multicolumn{11}{l}{\textbf{Model: Standard}} \\
        \midrule
        \multirow{3}{*}{High}
        & BBH  & $378.00^{+20.46}_{-19.43}$ & $343.00^{+19.54}_{-18.51}$ & $35.00^{+6.97}_{-5.89}$ & $99.00^{+10.98}_{-9.93}$ & $14.00^{+4.83}_{-3.70}$ & $0.00^{+1.84}_{-0.00}$ & $305.00^{+18.48}_{-17.46}$ & $15.00^{+4.96}_{-3.83}$ & $554.00^{+24.55}_{-23.53}$ \\
        & BNS  & $6.50^{+0.91}_{-0.80}$ & $4.80^{+0.80}_{-0.69}$ & $1.70^{+0.52}_{-0.41}$ & $1.90^{+0.54}_{-0.43}$ & $0.00^{+0.18}_{-0.00}$ & $0.00^{+0.18}_{-0.00}$ & $19.50^{+1.50}_{-1.40}$ & $2.40^{+0.60}_{-0.49}$ & $24.50^{+1.67}_{-1.56}$ \\
        & NSBH & $9.40^{+1.07}_{-0.97}$ & $8.90^{+1.05}_{-0.94}$ & $0.50^{+0.34}_{-0.22}$ & $3.10^{+0.66}_{-0.55}$ & $0.00^{+0.18}_{-0.00}$ & $0.00^{+0.18}_{-0.00}$ & $12.10^{+1.20}_{-1.10}$ & $0.50^{+0.34}_{-0.22}$ & $22.20^{+1.59}_{-1.49}$ \\
        \cmidrule(lr){1-11}
        \multirow{3}{*}{Low}
        & BBH  & $431.50^{+6.67}_{-6.57}$ & $395.40^{+6.39}_{-6.29}$ & $36.10^{+2.00}_{-1.90}$ & $107.10^{+3.37}_{-3.27}$ & $20.00^{+1.52}_{-1.41}$ & $0.40^{+0.32}_{-0.19}$ & $359.30^{+6.09}_{-5.99}$ & $15.90^{+1.36}_{-1.26}$ & $617.20^{+7.96}_{-7.86}$ \\
        & BNS  & $4.80^{+0.80}_{-0.69}$ & $3.60^{+0.71}_{-0.60}$ & $1.20^{+0.46}_{-0.34}$ & $1.60^{+0.51}_{-0.40}$ & $0.70^{+0.38}_{-0.26}$ & $0.00^{+0.18}_{-0.00}$ & $16.30^{+1.38}_{-1.28}$ & $2.00^{+0.55}_{-0.44}$ & $20.00^{+1.52}_{-1.41}$ \\
        & NSBH & $13.40^{+1.26}_{-1.16}$ & $12.50^{+1.22}_{-1.12}$ & $0.90^{+0.41}_{-0.29}$ & $3.10^{+0.66}_{-0.55}$ & $0.40^{+0.32}_{-0.19}$ & $0.00^{+0.18}_{-0.00}$ & $15.30^{+1.34}_{-1.24}$ & $0.40^{+0.32}_{-0.19}$ & $29.70^{+1.83}_{-1.72}$ \\
        \midrule

        \multicolumn{11}{l}{\textbf{Model: Optimistic CE}} \\
        \midrule
        \multirow{3}{*}{High}
        & BBH  & $1203.00^{+35.69}_{-34.68}$ & $1128.00^{+34.60}_{-33.58}$ & $75.00^{+9.70}_{-8.64}$ & $277.00^{+17.66}_{-16.63}$ & $81.00^{+10.04}_{-8.98}$ & $3.00^{+2.92}_{-1.63}$ & $827.00^{+29.77}_{-28.75}$ & $51.00^{+8.19}_{-7.12}$ & $1563.00^{+40.54}_{-39.53}$ \\
        & BNS  & $52.00^{+8.26}_{-7.19}$ & $47.00^{+7.90}_{-6.83}$ & $5.00^{+3.38}_{-2.16}$ & $14.00^{+4.83}_{-3.70}$ & $1.00^{+2.30}_{-0.83}$ & $0.00^{+1.84}_{-0.00}$ & $173.00^{+14.18}_{-13.14}$ & $12.00^{+4.56}_{-3.42}$ & $237.00^{+16.42}_{-15.38}$ \\
        & NSBH & $18.70^{+1.47}_{-1.37}$ & $17.10^{+1.41}_{-1.31}$ & $1.60^{+0.51}_{-0.40}$ & $4.10^{+0.75}_{-0.64}$ & $0.60^{+0.36}_{-0.24}$ & $0.00^{+0.18}_{-0.00}$ & $22.40^{+1.60}_{-1.50}$ & $2.00^{+0.55}_{-0.44}$ & $40.80^{+2.12}_{-2.02}$ \\
        \cmidrule(lr){1-11}
        \multirow{3}{*}{Low}
        & BBH  & $990.00^{+32.48}_{-31.46}$ & $920.00^{+31.34}_{-30.33}$ & $70.00^{+9.41}_{-8.35}$ & $249.00^{+16.80}_{-15.77}$ & $46.00^{+7.83}_{-6.76}$ & $4.00^{+3.16}_{-1.91}$ & $739.00^{+28.20}_{-27.18}$ & $41.00^{+7.46}_{-6.38}$ & $1351.00^{+37.77}_{-36.75}$ \\
        & BNS  & $60.00^{+8.79}_{-7.72}$ & $45.00^{+7.76}_{-6.68}$ & $15.00^{+4.96}_{-3.83}$ & $21.00^{+5.66}_{-4.55}$ & $2.00^{+2.64}_{-1.29}$ & $1.00^{+2.30}_{-0.83}$ & $172.00^{+14.14}_{-13.10}$ & $15.00^{+4.96}_{-3.83}$ & $245.00^{+16.67}_{-15.64}$ \\
        & NSBH & $19.40^{+1.50}_{-1.39}$ & $17.90^{+1.44}_{-1.34}$ & $1.50^{+0.50}_{-0.38}$ & $5.40^{+0.84}_{-0.73}$ & $0.70^{+0.38}_{-0.26}$ & $0.00^{+0.18}_{-0.00}$ & $23.70^{+1.64}_{-1.54}$ & $1.30^{+0.47}_{-0.36}$ & $43.30^{+2.18}_{-2.08}$ \\
        \midrule

        \multicolumn{11}{l}{\textbf{Model: Delayed SN}} \\
        \midrule
        \multirow{3}{*}{High}
        & BBH  & $395.00^{+20.89}_{-19.87}$ & $352.00^{+19.78}_{-18.75}$ & $43.00^{+7.61}_{-6.53}$ & $80.00^{+9.98}_{-8.93}$ & $10.00^{+4.27}_{-3.11}$ & $0.00^{+1.84}_{-0.00}$ & $324.00^{+19.02}_{-17.99}$ & $16.00^{+5.08}_{-3.96}$ & $556.00^{+24.59}_{-23.57}$ \\
        & BNS  & $6.20^{+1.33}_{-1.11}$ & $4.60^{+1.17}_{-0.95}$ & $1.60^{+0.79}_{-0.55}$ & $1.80^{+0.82}_{-0.59}$ & $0.20^{+0.46}_{-0.17}$ & $0.00^{+0.37}_{-0.00}$ & $20.60^{+2.24}_{-2.03}$ & $1.80^{+0.82}_{-0.59}$ & $24.20^{+2.41}_{-2.20}$ \\
        & NSBH & $5.10^{+0.82}_{-0.71}$ & $4.50^{+0.78}_{-0.67}$ & $0.60^{+0.36}_{-0.24}$ & $1.10^{+0.44}_{-0.33}$ & $0.10^{+0.23}_{-0.08}$ & $0.10^{+0.23}_{-0.08}$ & $7.30^{+0.96}_{-0.85}$ & $0.80^{+0.39}_{-0.28}$ & $12.00^{+1.20}_{-1.09}$ \\
        \cmidrule(lr){1-11}
        \multirow{3}{*}{Low}
        & BBH  & $284.00^{+17.87}_{-16.84}$ & $260.00^{+17.15}_{-16.11}$ & $24.00^{+5.97}_{-4.86}$ & $77.00^{+9.81}_{-8.76}$ & $14.00^{+4.83}_{-3.70}$ & $1.00^{+2.30}_{-0.83}$ & $247.00^{+16.74}_{-15.71}$ & $11.00^{+4.42}_{-3.27}$ & $431.00^{+21.78}_{-20.75}$ \\
        & BNS  & $8.00^{+1.48}_{-1.26}$ & $7.40^{+1.43}_{-1.21}$ & $0.60^{+0.58}_{-0.33}$ & $1.80^{+0.82}_{-0.59}$ & $0.00^{+0.37}_{-0.00}$ & $0.00^{+0.37}_{-0.00}$ & $18.00^{+2.10}_{-1.89}$ & $1.80^{+0.82}_{-0.59}$ & $25.40^{+2.46}_{-2.25}$ \\
        & NSBH & $5.00^{+0.81}_{-0.70}$ & $5.00^{+0.81}_{-0.70}$ & $0.30^{+0.29}_{-0.16}$ & $1.00^{+0.43}_{-0.31}$ & $0.00^{+0.18}_{-0.00}$ & $0.00^{+0.18}_{-0.00}$ & $5.30^{+0.83}_{-0.73}$ & $0.20^{+0.26}_{-0.13}$ & $13.30^{+1.26}_{-1.15}$ \\
        \midrule

        \multicolumn{11}{l}{\textbf{Model: High BH Kicks}} \\
        \midrule
        \multirow{3}{*}{High}
        & BBH  & $31.00^{+6.63}_{-5.54}$ & $27.00^{+6.26}_{-5.16}$ & $4.00^{+3.16}_{-1.91}$ & $7.00^{+3.77}_{-2.58}$ & $0.00^{+1.84}_{-0.00}$ & $0.00^{+1.84}_{-0.00}$ & $23.00^{+5.87}_{-4.76}$ & $1.00^{+2.30}_{-0.83}$ & $48.00^{+7.98}_{-6.90}$ \\
        & BNS  & $5.20^{+1.23}_{-1.01}$ & $4.80^{+1.19}_{-0.97}$ & $0.40^{+0.53}_{-0.26}$ & $1.60^{+0.79}_{-0.55}$ & $0.60^{+0.58}_{-0.33}$ & $0.00^{+0.37}_{-0.00}$ & $16.00^{+2.00}_{-1.79}$ & $2.00^{+0.85}_{-0.62}$ & $23.00^{+2.35}_{-2.14}$ \\
        & NSBH & $1.30^{+0.47}_{-0.36}$ & $1.20^{+0.46}_{-0.34}$ & $0.10^{+0.23}_{-0.08}$ & $0.40^{+0.32}_{-0.19}$ & $0.00^{+0.18}_{-0.00}$ & $0.00^{+0.18}_{-0.00}$ & $1.80^{+0.53}_{-0.42}$ & $0.00^{+0.18}_{-0.00}$ & $3.30^{+0.68}_{-0.57}$ \\
        \cmidrule(lr){1-11}
        \multirow{3}{*}{Low}
        & BBH  & $18.00^{+5.32}_{-4.20}$ & $18.00^{+5.32}_{-4.20}$ & $0.00^{+1.84}_{-0.00}$ & $4.00^{+3.16}_{-1.91}$ & $1.00^{+2.30}_{-0.83}$ & $0.00^{+1.84}_{-0.00}$ & $21.00^{+5.66}_{-4.55}$ & $2.00^{+2.64}_{-1.29}$ & $32.00^{+6.72}_{-5.63}$ \\
        & BNS  & $7.40^{+1.43}_{-1.21}$ & $5.60^{+1.27}_{-1.05}$ & $1.80^{+0.82}_{-0.59}$ & $2.20^{+0.88}_{-0.65}$ & $0.20^{+0.46}_{-0.17}$ & $0.00^{+0.37}_{-0.00}$ & $17.60^{+2.08}_{-1.87}$ & $0.60^{+0.58}_{-0.33}$ & $26.00^{+2.49}_{-2.28}$ \\
        & NSBH & $0.90^{+0.41}_{-0.29}$ & $0.90^{+0.41}_{-0.29}$ & $0.00^{+0.18}_{-0.00}$ & $0.20^{+0.26}_{-0.13}$ & $0.00^{+0.18}_{-0.00}$ & $0.00^{+0.18}_{-0.00}$ & $1.00^{+0.43}_{-0.31}$ & $0.00^{+0.18}_{-0.00}$ & $2.20^{+0.58}_{-0.47}$ \\
        \bottomrule
        \bottomrule
    \end{tabular*}

    \vspace{0.5em}
    \begin{minipage}{\textwidth}
        \footnotesize
        \textbf{Note:}
        This table presents a hierarchical summary of the predicted annual lensed GW event rates for the combined ET+CE detector network. The results are stratified by four distinct binary evolution models (Standard, Optimistic CE, Delayed SN, and High BH Kicks), two metallicity evolution scenarios (High and Low), and three compact binary source types (BBH, BNS, and NSBH). The columns quantify events across different lensing morphologies: $N_{\mathrm{multi}}$ aggregates double ($N_{\mathrm{double}}$) and quadruple ($N_{\mathrm{quad}}$) image systems; $N_{\mathrm{sub}}$ denotes events dominated by subhalo lensing; and $N^{\mathrm{CI}}_{3,5}$ represents the system capable of detecting the central image. $N_{\text{any}>8}$ represents the total count of systems with at least one image having an SNR $> 8$. High-magnification events are categorized by maximum magnification thresholds of $|\mu| > 3$ and $|\mu| > 10$. Uncertainties correspond to the 68\% confidence level.
    \end{minipage}
     \label{tab:lensed_events_models_full_comparison}
\end{table*}

To investigate the sensitivity of lensed GW event rates to astrophysical modeling assumptions, we performed a comparative analysis of four isolated binary evolution models and two distinct mass models. These comparisons are based on a one-year observation period and assume the joint operation of the ET and CE detector network.

\begin{table*}[t]
    \centering
    \caption{Lensed GW Event Rates ($\mathrm{yr}^{-1}$) Arranged by Image Classification for Different Mass Models}
    \label{tab:mass_model}

    \scriptsize
    \renewcommand{\arraystretch}{1.5}
    \setlength{\tabcolsep}{0pt}

    \begin{tabular*}{\textwidth}{@{\extracolsep{\fill}}cccccccccc}
        \toprule
        \toprule

        \multirow{2}{*}{\textbf{Mass Model}} &
        \multicolumn{4}{c}{\textbf{Multiple images ($\mathrm{yr}^{-1}$)}} &
        \multicolumn{2}{c}{\textbf{Central Image cases ($\mathrm{yr}^{-1}$)}} &
        \multicolumn{2}{c}{\textbf{High-mag ($\mathrm{yr}^{-1}$)}} &
        \multirow{2}{*}{\textbf{$N_{\mathrm{trig}}$}} \\

        \cmidrule(lr){2-5} \cmidrule(lr){6-7} \cmidrule(lr){8-9}

        & $N_{\mathrm{multi}}$ & $N_{\mathrm{double}}$ & $N_{\mathrm{quad}}$ & $N_{\mathrm{sub}}$
        & $N^{\mathrm{CI}}_{3}$ & $N^{\mathrm{CI}}_{5}$
        & $N_{|\mu|>3}$ & $N_{|\mu|>10}$
        & (Any $> 8$) \\
        \midrule

        \multicolumn{10}{l}{\textbf{\textit{Broken Power Law + 2 Peaks}}} \\
        \midrule
        \hspace{1em} $5\text{--}50\,M_\odot$ &
        $438.00^{+21.94}_{-20.92}$ & $404.00^{+21.12}_{-20.09}$ & $34.00^{+6.89}_{-5.80}$ & $113.00^{+11.66}_{-10.61}$ &
        $24.00^{+5.97}_{-4.86}$ & $0.00^{+1.84}_{-0.00}$ &
        $373.00^{+20.33}_{-19.30}$ & $18.00^{+5.32}_{-4.20}$ &
        $626.00^{+26.03}_{-25.01}$ \\

        \hspace{1em} $5\text{--}85\,M_\odot$ &
        $447.00^{+22.16}_{-21.13}$ & $409.00^{+21.24}_{-20.22}$ & $38.00^{+7.22}_{-6.14}$ & $111.00^{+11.57}_{-10.52}$ &
        $18.00^{+5.32}_{-4.20}$ & $1.00^{+2.30}_{-0.83}$ &
        $383.00^{+20.59}_{-19.56}$ & $25.00^{+6.07}_{-4.97}$ &
        $634.00^{+26.19}_{-25.17}$ \\

        \hspace{1em} $3\text{--}300\,M_\odot$ &
        $431.50^{+6.67}_{-6.57}$ & $395.40^{+6.39}_{-6.29}$ & $36.10^{+2.00}_{-1.90}$ & $107.10^{+3.37}_{-3.27}$ &
        $20.00^{+1.52}_{-1.41}$ & $0.40^{+0.32}_{-0.19}$ &
        $359.30^{+6.09}_{-5.99}$ & $15.90^{+1.36}_{-1.26}$ &
        $617.20^{+7.96}_{-7.86}$ \\
        \midrule

        \multicolumn{10}{l}{\textbf{\textit{Power Law}}} \\
        \midrule
        \hspace{1em} $\alpha=-2.5$ ($5\text{--}50\,M_\odot$) &
        $404.00^{+21.12}_{-20.09}$ & $368.00^{+20.20}_{-19.18}$ & $36.00^{+7.06}_{-5.97}$ & $102.00^{+11.13}_{-10.08}$ &
        $18.00^{+5.32}_{-4.20}$ & $1.00^{+2.30}_{-0.83}$ &
        $321.00^{+18.94}_{-17.91}$ & $16.00^{+5.08}_{-3.96}$ &
        $592.00^{+25.35}_{-24.32}$ \\

        \hspace{1em} $\alpha=-1.0$ ($5\text{--}85\,M_\odot$) &
        $495.00^{+23.26}_{-22.24}$ & $451.00^{+22.25}_{-21.23}$ & $44.00^{+7.68}_{-6.61}$ & $105.00^{+11.28}_{-10.23}$ &
        $32.00^{+6.72}_{-5.63}$ & $2.00^{+2.64}_{-1.29}$ &
        $352.00^{+19.78}_{-18.75}$ & $14.00^{+4.83}_{-3.70}$ &
        $603.00^{+25.57}_{-24.55}$ \\
        \bottomrule
        \bottomrule
    \end{tabular*}

    \vspace{0.5em}
    \begin{minipage}{\textwidth}
        \footnotesize
        \textbf{Note:}
        This table presents a comparative analysis of predicted annual lensed GW event rates under different primary mass distribution models. We evaluate the standard Broken Power Law + 2 Peaks model \citep{theligoscientificcollaboration2025gwtc40populationpropertiesmerging} across three mass ranges ($5\text{--}50\,M_\odot$, $5\text{--}85\,M_\odot$, and the fiducial $3\text{--}300\,M_\odot$) and compare it with simple Power Law models defined by different spectral indices ($\alpha$) and mass cutoffs.
        Columns are categorized by image morphology: $N_{\mathrm{multi}}$ aggregates double ($N_{\mathrm{double}}$) and quadruple ($N_{\mathrm{quad}}$) images, with subhalo-lensed events ($N_{\mathrm{sub}}$) listed separately.
        $N^{\mathrm{CI}}_{3,5}$ represents the system capable of detecting the central image.
        High-magnification events are counted if $|\mu| > 3$ or $> 10$.
        $N_{\mathrm{trig}}$ represents the total count of triggered events (at least one image with SNR $> 8$).
        All rates are calculated for the ET+CE network. Uncertainties correspond to the 68\% confidence level.
    \end{minipage}
\end{table*}

\subsubsection{Impact of Binary Evolution Models}

We performed a comparative analysis of lensed GW event rates across four distinct binary evolution models: the Standard Model, the Optimistic CE model, the Delayed SN model, and the High BH Kicks model according to \cite{2013ApJ...779...72D}. These evaluations were conducted under both high and low metallicity evolution scenarios, with the results summarized in \autoref{tab:lensed_events_models_full_comparison}.

The data indicate that the Optimistic CE model yields approximately three times the total number of GW events compared to the Standard model, consequently resulting in the highest number of lensed events. In contrast, the High BH Kicks model produces only about $1/20$ of the total event count found in the Standard model, leading to the fewest lensed events. This demonstrates that the total number of intrinsic GW events is the primary driver of the lensed event population. Consequently, variations in total event counts across different merger rate models significantly impact our assessment of the overall lensed GW population, suggesting that the detection prospects for lensed GWs are highly dependent on the underlying physics of stellar formation and binary evolution.

Compared to high-metallicity environments, low-metallicity scenarios generally yield higher event rates. This is attributed to the fact that low-metallicity environments are typically more favorable to the formation of massive black holes and exhibit higher binary merger efficiencies. Notably, however, the impact of metallicity variations is substantially smaller than that of the binary evolution model selection, indicating that uncertainties in evolutionary pathways dominate the overall systematic error.

\subsubsection{Sensitivity to Mass Models}
Furthermore, we performed a comparative analysis with the mass models adopted in \citet{10.1093/mnras/stab3298}. Our primary objective was to evaluate the impact of the \citet{theligoscientificcollaboration2025gwtc40populationpropertiesmerging} baseline model (Broken Power Law + 2 Peaks) versus the Power Law mass model on the joint detection rates of the ET and CE network across different mass ranges (see \autoref{tab:mass_model}). Initially, by modifying only the mass model while maintaining all other parameters consistent with that study, we adopted a Power Law model with a spectral index of $\alpha = 2.3$ within the mass range of $5$--$50\,M_\odot$. Under this configuration, we derived an ET+CE event rate of $273.00^{+17.54}_{-16.51}$, which represents a decrease of approximately 20\% compared to their prediction. Similarly, for a Power Law model with $\alpha = -1$ spanning a mass range of $5$--$85\,M_\odot$, our predicted event rate for ET+CE is 18\% lower than that reported therein. Moreover, under the composite lens mass model adopted in this study, the predicted increasing trend in the lensed GW event rates for the ET+CE network across the two Power Law models aligns well with previous findings (22\%; this work: 24\%).

A comparative analysis of predicted lensed GW event rates using the Broken Power Law + 2 Peaks model and the Power Law model reveals a distinct contrast in their sensitivity to the defined mass range. The Power Law model proves to be highly sensitive to the mass cutoff, whereas the Broken Power Law + 2 Peaks model exhibits remarkable stability, with event rate fluctuations generally confined to within 10 events even as the mass range is expanded. {We note that the Power Law + peak model yields results consistent with the Broken Power Law + 2 Peaks configuration used here. Our comparison primarily focuses on the latter to illustrate the divergence from simple power-law models commonly used in earlier literature.}

This stability is attributed to the intrinsic structure of the BBH mass spectrum, which declines significantly beyond the second peak at $35\,M_\odot$, rendering systems with masses exceeding $50\,M_\odot$ scarce. Indeed, simulations based on the \citet{theligoscientificcollaboration2025gwtc40populationpropertiesmerging} benchmark model indicate that approximately 99\% of the BBH population is concentrated within the $5$--$50\,M_\odot$ range. Given that higher-mass BBH events intrinsically yield higher SNRs, a simple power-law distribution tends to overestimate the lensed event rate by artificially extending the high-mass tail. Consequently, the observed insensitivity of the Broken Power Law + 2 Peaks model to mass range variations is consistent with the underlying physical constraints of the population.

\subsection{Distributions}
Based on ten years of simulated GW observations utilizing the combined ET and CE network, we analyze the parameter distributions for both sources and lenses.
Our discussion primarily focuses on the redshift distributions of sources and lenses, the impact of magnification bias on source redshift, the distribution of maximum angular separations in multi-image events, as well as the distributions of SNR and time delays.
All respective parameters of the source, the lens system and images have been released in the \texttt{GW-LMC} mock catalog on GitHub \footnote{\url{https://github.com/LensedGW/GW-LMC}}and have been archived on Zenodo\footnote{\url{https://zenodo.org/records/19212271}} \citep{li_youkai_2026_19212271}.

\subsubsection{Redshift and Angular Separation Distributions}
\begin{figure}[t!]
    \centering
    \hspace*{-0.5cm}
    \includegraphics[width=1.1\linewidth]{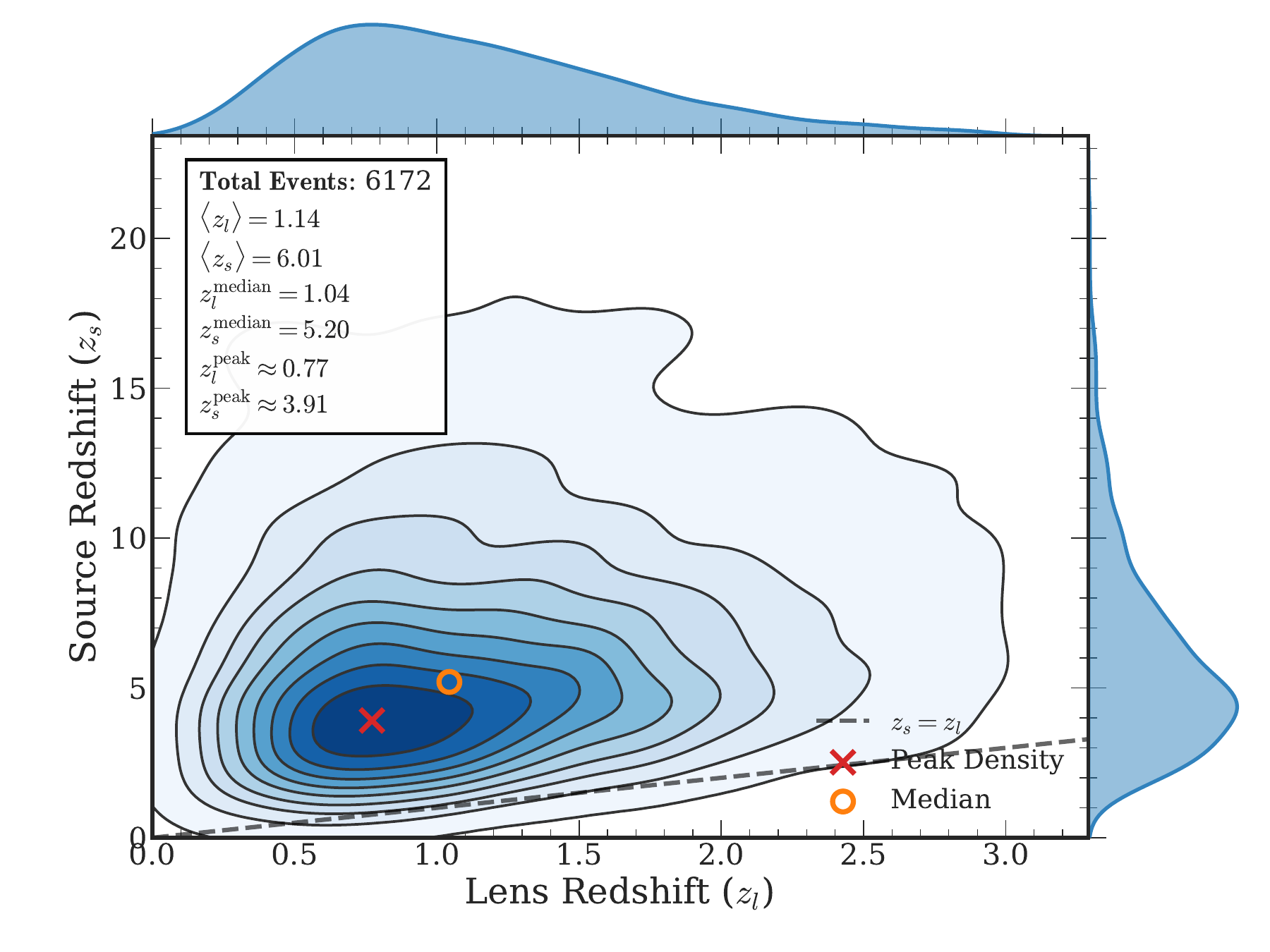}
    \caption{The joint probability density distribution of the lens redshift ($z_l$) and source redshift ($z_s$) for 6,172 simulated strongly lensed GW events from the $N_{\mathrm{trig}}$ catalog. The filled curves on the top and right margins illustrate the marginal distributions for $z_l$ and $z_s$, respectively. Within the joint contour plot, the red cross marks the location of the peak probability density, while the orange circle represents the median of the distribution. The diagonal dashed gray line indicates the $z_s = z_l$ reference line.}
    \label{fig:z distribution}
\end{figure}

Utilizing the complete $N_{\mathrm{trig}}$ catalog, which comprises 6,172 strongly lensed multi-image events, we investigate the joint distribution of the source redshift, $z_s$, and the lens redshift, $z_l$ (see \autoref{fig:z distribution}).
Statistical analysis yields a mean source redshift of $\langle z_s \rangle = 6.01$ and a mean lens redshift of $\langle z_l \rangle = 1.14$.
However, given the significant heavy-tailed nature of the distribution, we also extract the median and peak values to more robustly characterize the most typical event topology.
Specifically, the medians for the source and lens redshifts are $z_s^{\mathrm{median}} = 5.20$ and $z_l^{\mathrm{median}} = 1.04$, respectively, while the probability density peaks at $z_s^{\mathrm{peak}} \approx 3.91$ and $z_l^{\mathrm{peak}} \approx 0.77$.
This indicates that the bulk of the lensed GW population originates from sources at $z_s \sim 3.9$ and is lensed by galaxies at $z_l \sim 0.8$.

We further investigate the impact of magnification bias on the redshift of GW sources.
A primary scientific objective of third-generation GW detectors is the detection of high-redshift signals.
In this context, the magnification bias induced by gravitational lensing serves as a crucial mechanism for extending our observational horizon to higher redshifts.
We define ``high-magnification events'' as those with a maximum image magnification of $\mu_{\mathrm{max}} > 3$ in the lens system; this category encompasses not only multiple-image systems but also highly magnified single-image events.

\begin{figure}[t!]
    \centering
    \includegraphics[width=1\linewidth]{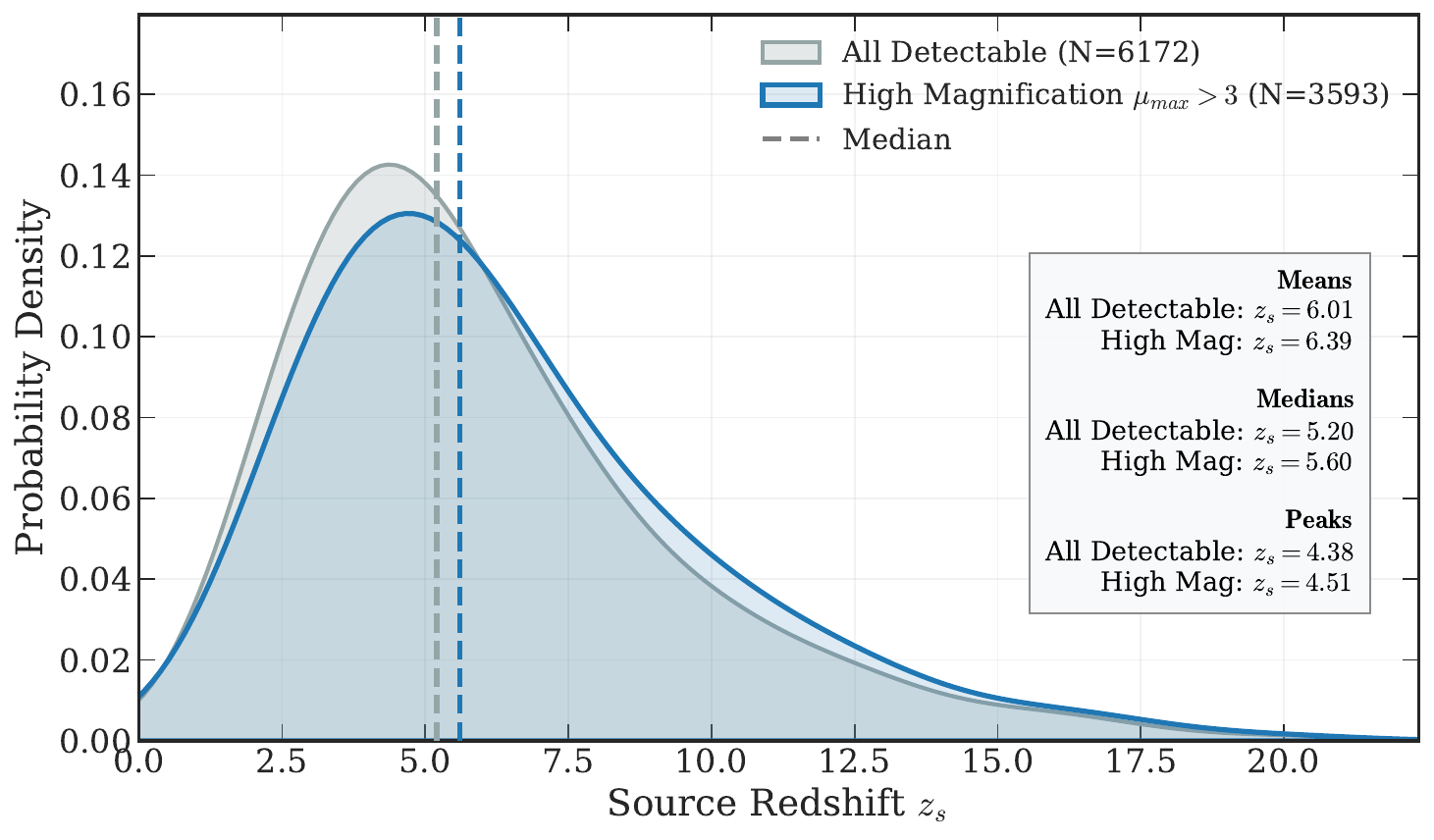}
    \caption{Impact of magnification bias on the redshift distribution of observable lensed GW events. The distribution for all detectable events ($N=6172$) peaks at $z_s \approx 4.38$ with a median of $z_s = 5.20$, whereas high-magnification events ($\mu_{\mathrm{max}} > 3$, including both single and multiple images; $N=3593$) exhibit a peak shifted to a higher redshift of $z_s \approx 4.51$ and a median of $z_s = 5.60$. Both distributions feature an extended tail reaching $z_s > 15$, demonstrating that high-magnification lensing events serve as unique probes for binary mergers in the early Universe.}
    \label{fig:mag_bias}
\end{figure}

In \autoref{fig:mag_bias}, the gray-shaded region represents the population of all detectable multi-image lensing events ($N=6172$), characterized by a marginal source redshift distribution that peaks at $z_s \approx 4.38$ with a median of $z_s = 5.20$.
In contrast, the high-magnification sub-population ($N=3593$) exhibits a peak shifted to $z_s \approx 4.51$, with its median correspondingly increasing to $z_s = 5.60$.
Furthermore, both distributions display significant heavy-tailed features, extending into the extremely high-redshift regime of $z_s > 15$ and even reaching $z_s \sim 20$.
Consequently, focusing on high-magnification lensed GWs offers a unique opportunity to probe the formation and evolution of binary mergers in the early Universe.

Using the simulated data, we further investigate the distribution of the maximum angular separation ($\Delta \theta$) for lensed GW signals. It is important to emphasize that, given the typically poor sky localization of GW detectors, individual lensed images separated by mere arcseconds cannot be spatially resolved. Nevertheless, characterizing the intrinsic $\Delta \theta$ distribution remains crucial in this context. Primarily, it provides essential structural parameters for multi-messenger astronomy, guiding high-resolution electromagnetic follow-up observations aimed at identifying the host lens galaxy or potential counterparts. Furthermore, the angular separation is fundamentally tied to the underlying lens mass distribution and the resulting time delays between images.

\begin{figure}[t!]
    \centering
    \includegraphics[width=1\linewidth]{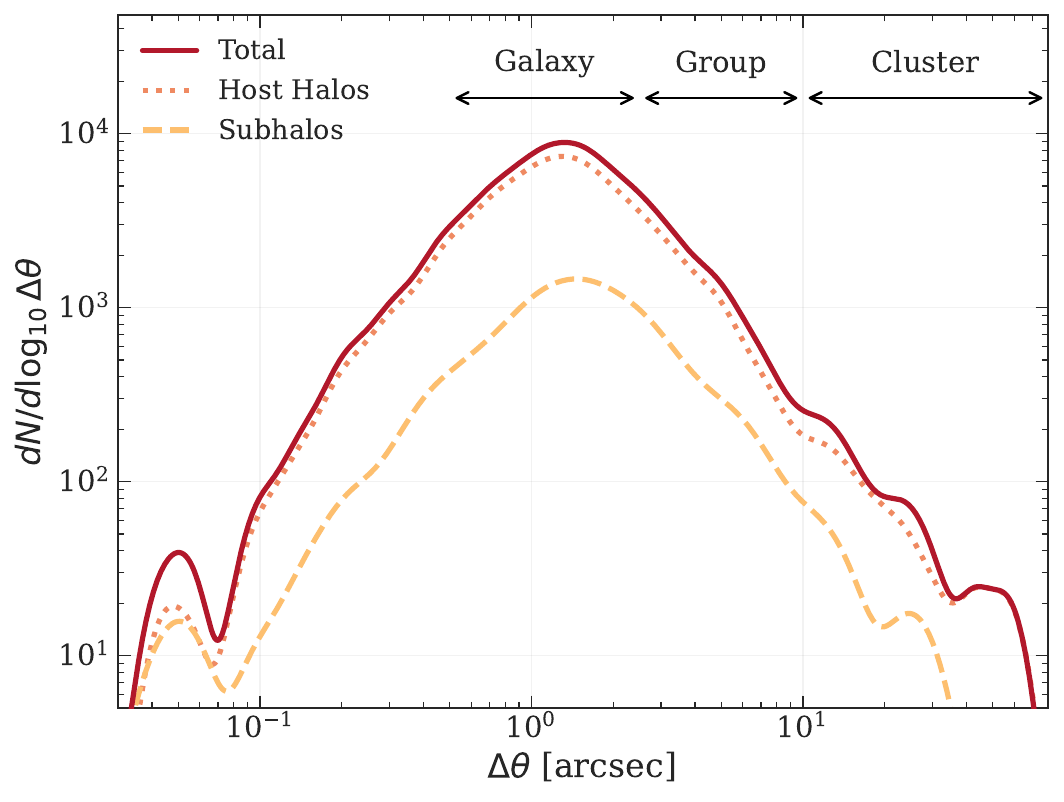}
    \caption{Differential distribution of the maximum angular separation ($\Delta \theta$) between signal copies for lensed GW multi-image events, plotted as $dN/d\log_{10} \Delta\theta$. The distribution encompasses three distinct lens mass scales: galaxy, group, and cluster scales. Furthermore, contributions from different lens structures are distinguished: host halos are indicated by the red dotted line, subhalos by the orange dashed line, and the total population (host + subhalo) by the red solid line. The overall distribution is predominantly dominated by galaxy-scale lenses, peaking at $\Delta \theta \approx 1.28$ arcseconds, while the cluster scale constitutes the extended tail at large angular separations.}
    \label{fig:angular}
\end{figure}

As illustrated in \autoref{fig:angular}, the distribution is predominantly governed by galaxy-scale lenses. Quantitatively, approximately $83.5\%$ of the lensing events fall within the galaxy scale, while the group and cluster scales contribute roughly $15\%$ and $1.5\%$, respectively. Regarding lens structures, host halos dominate the population ($75\%$), with subhalos contributing approximately $25\%$. The distribution peaks at $\Delta \theta \approx 1.28$ arcseconds, which is consistent with the typical Einstein radii of massive galactic lenses. Notably, the distribution morphology of subhalos closely traces that of the host halos, suggesting similar statistical characteristics in the angular scales of their lensing effects. In addition, a secondary feature is observed in the small-scale regime ($\Delta \theta < 0.1$ arcseconds); although these events occur less frequently, they may correspond to extremely short time delays or regimes where wave optics effects become significant. Conversely, the cluster scale ($>10$ arcseconds) exhibits an extended tail, representing rare, high-mass lensing events capable of producing substantial time delays.
\begin{figure}[htbp]
    \centering
    \includegraphics[width=1\linewidth]{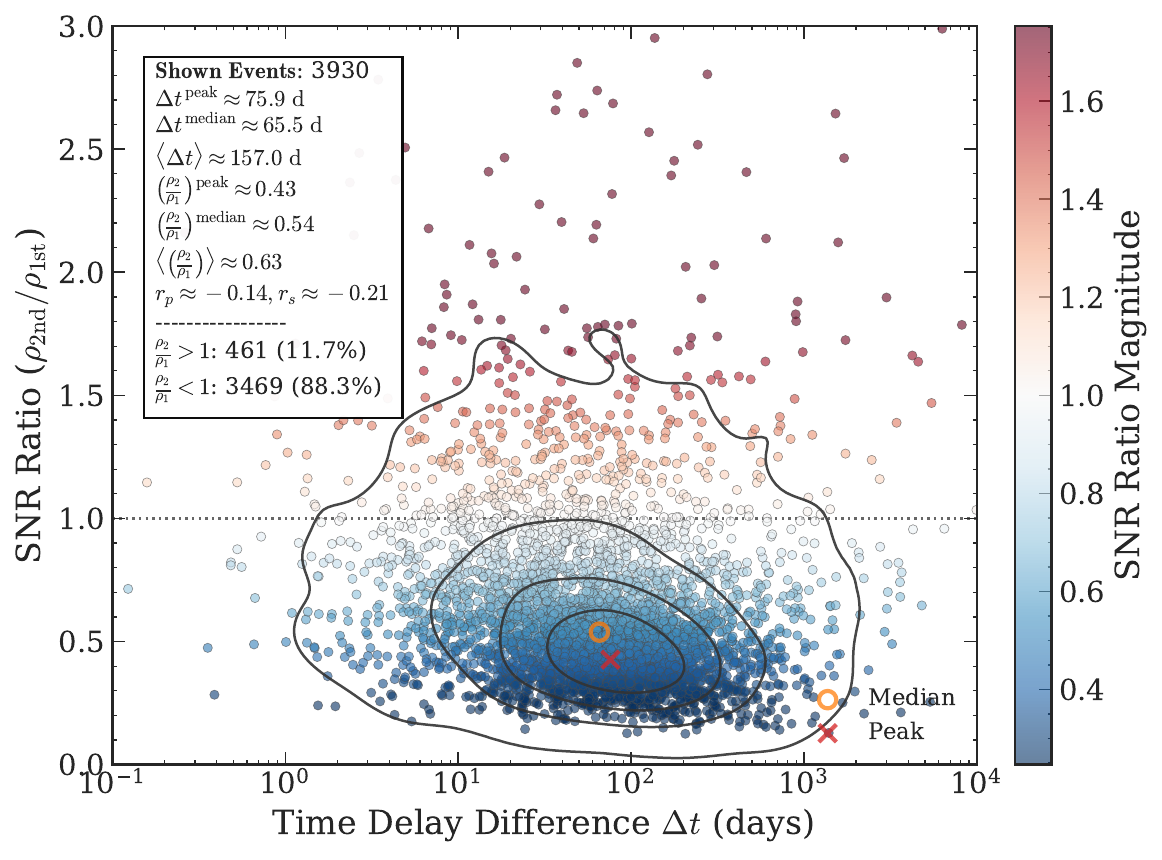}
    \caption{Distribution of the SNR ratio ($\rho_2/\rho_1$) versus time delay ($\Delta t$) for a sample of 3,930 double-image systems. The scatter plot illustrates the relationship between relative signal strength and time delay, color-coded by the magnitude of the SNR ratio. The overlaid black curve represents the Kernel Density Estimation (KDE) fit of the distribution. Key statistical metrics, including the peak value, mean, median, and Pearson and Spearman correlation coefficients, are provided in the legend. A horizontal dashed line at $\rho_2/\rho_1 = 1$ delineates the boundary between events where the second image is magnified ($11.7\%$) versus demagnified ($88.3\%$) relative to the first.}
    \label{fig:double_td_snr_ratio}
\end{figure}
\subsubsection{SNR Ratio and Time Delay Distribution}
To facilitate the identification of lensed GW events in future detectors, we investigate the relationship between the SNR ratio and the time delay of multiple-image events. We first focus our analysis on double-image events (see \autoref{fig:double_td_snr_ratio}), selecting a sample of 3,930 lensed GW doublets with time delays ranging from 0.1 to 10,000 days. We observe that in the vast majority of these double-image systems ($88.3\%$), the SNR of the leading signal exceeds that of the trailing signal ($\rho_2 / \rho_1 < 1$). The distribution of $\rho_2 / \rho_1$ peaks at $0.43$, with a median of $0.54$ and a mean of $0.63$, indicating that the second signal is typically demagnified relative to the first. Consequently, for most lens configurations, if $\rho_1$ marginally exceeds the detection threshold, $\rho_2$ is highly likely to fall below the detector's sensitivity limit.

\begin{figure*}[t!]
    \centering
    \includegraphics[width=1\textwidth]{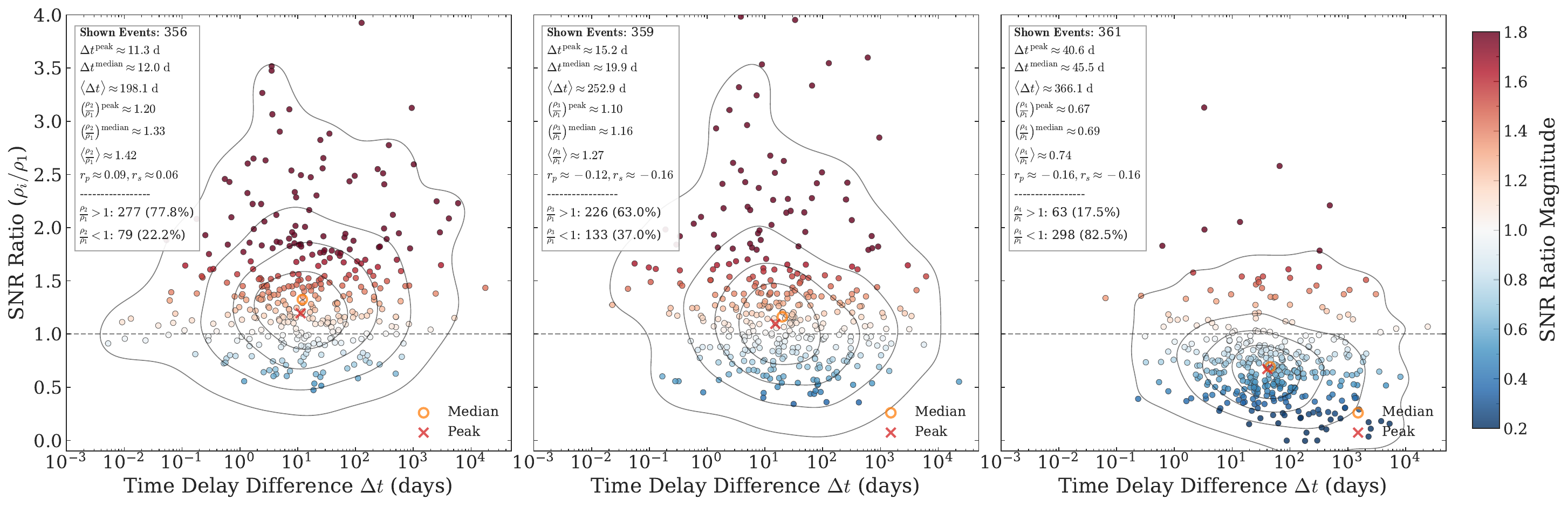}
    \caption{Comparative analysis of the SNR ratio ($\rho_i/\rho_1$) and time delay ($\Delta t$) for quadruple-image lensed GW systems. The three panels display the relationship between the first arriving image ($\rho_1$) and the subsequent second ($\rho_2$, left), third ($\rho_3$, center), and fourth ($\rho_4$, right) images. Each subplot features a scatter plot of the data overlaid with a KDE fit of the distribution. The legends summarize key statistics, including the peak, mean, and median values for both the SNR ratios and time delays, correlation coefficients ($r_p, r_s$), and the proportions of events where the subsequent image is louder ($\rho_i/\rho_1 > 1$) or fainter ($\rho_i/\rho_1 < 1$) than the first. Note the distinct behavior of the second and third images, which are frequently magnified relative to the first, in contrast to the typically demagnified fourth image.}
    \label{fig:quad_snr_td}
\end{figure*}

In the time domain, the time delays between images span a broad range, from several hours to several years. The distribution exhibits a peak at $75.9$ days and a median of $65.5$ days, whereas the mean time delay is approximately $157.0$ days. This discrepancy indicates that the distribution is significantly skewed, with its mean being heavily influenced by a small subset of events with exceptionally large time delays. Furthermore, correlation analysis reveals a weak negative correlation between the SNR ratio and the lensing time delay ($r_p \approx -0.14$, $r_s \approx -0.21$). This behavior can likely be attributed to the lens geometry: sources located closer to the lens center tend to produce signals with shorter time delays and magnification ratios closer to unity.

As illustrated in \autoref{fig:quad_snr_td}, the analysis of quadruple-image systems reveals characteristics that are distinct from those of double-image events. Following the convention established for double events, we define the first signal arriving at Earth as the primary image of the lens system. In contrast to the prevalent demagnification observed in double events, quad events exhibit a distinct trend where late-arriving signals frequently possess higher SNR ratios.

Specifically, the subplots for the $\rho_2/\rho_1$ and $\rho_3/\rho_1$ distributions indicate that in $77.8\%$ and $63.0\%$ of the cases, respectively, the late-arriving signals satisfy $\rho_{\mathrm{late}} / \rho_{\mathrm{early}} > 1$. The median SNR ratios for these two subsequent images reach $1.33$ and $1.16$ (with peaks at $1.20$ and $1.10$), respectively. Furthermore, their time delay distributions are notably similar, with median time delays of $12.0$ days and $19.9$ days for the second and third images, respectively. Physically, this phenomenon can be attributed to the formation of highly magnified image pairs when the source is positioned near the caustics of the lens, implying that the initial trigger is frequently not the most luminous signal.

Conversely, the fourth signal ($\rho_4$) undergoes demagnification in the vast majority of instances ($82.5\%$), with a median SNR ratio of $0.69$. It is also typically accompanied by substantial time delays, exhibiting a median time delay of $45.5$ days (and a remarkably high mean of $366.1$ days). These statistical insights impose novel requirements on multi-messenger and follow-up observation strategies: for quadruple lens candidates, the peak sensitivity signal often resides in the middle of the arrival sequence. Consequently, upon identifying a potential lensed trigger, it is imperative not only to monitor for subsequent faint signals (relevant to $\rho_4$) but also to perform retrospective searches of archival data. Such ``pre-trigger'' searches are crucial for recovering earlier, potentially sub-threshold pioneer signals ($\rho_1$) that may have been initially overlooked.

\begin{figure}[t!]
    \centering
    \includegraphics[width=1\linewidth]{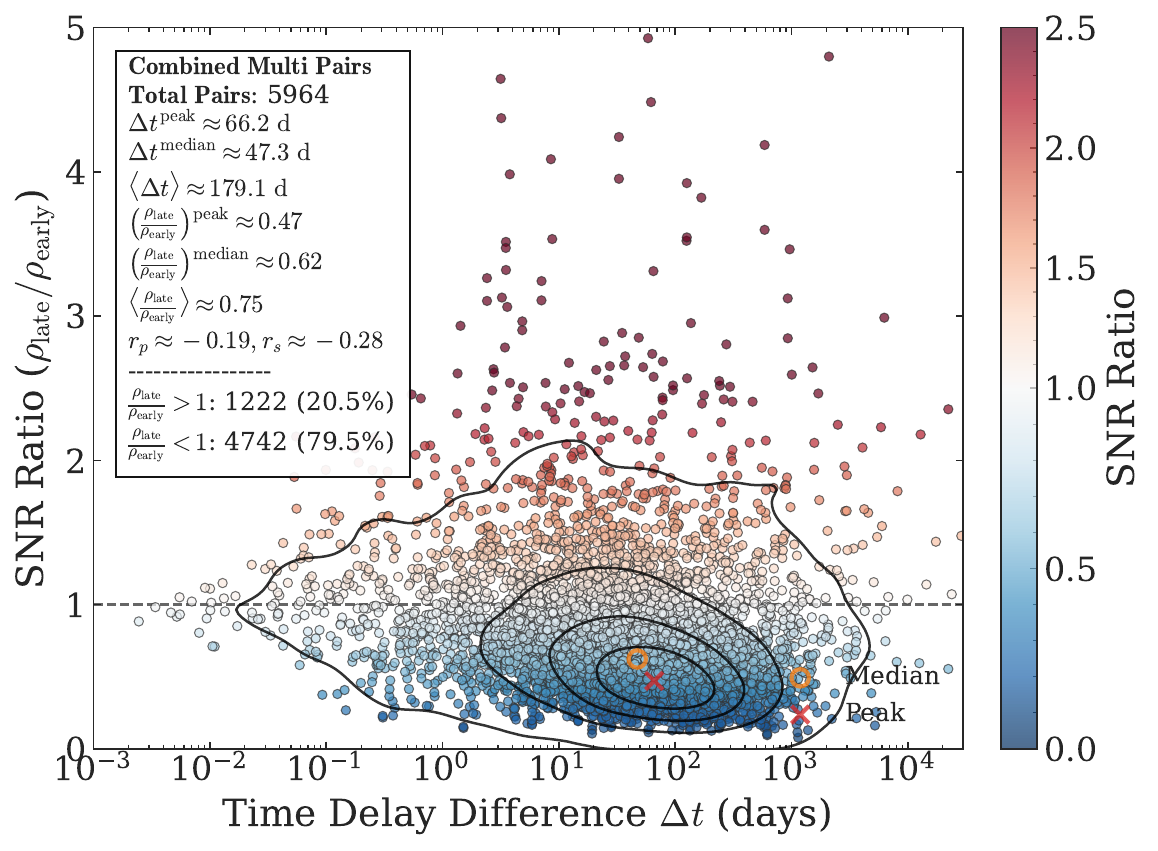}
    \caption{Joint distribution of the SNR ratio ($\rho_{\mathrm{late}}/\rho_{\mathrm{early}}$) and time delay ($\Delta t$) for the combined population of all detected multiple-image GW events within the $N_{\mathrm{trig}}$ dataset (totaling 5,964 pairs). For quadruple systems, all possible combinatorial pairs among detectable images are included. The main scatter plot is color-coded by the SNR ratio, with the black solid line representing the KDE fit of the distribution. The legend provides the peak, mean, median, and correlation coefficients, along with the percentage of cases where the late-arriving image is more luminous than the early-arriving one (approximately $20.5\%$).}
    \label{fig:multi_snr_td}
\end{figure}

From an observational perspective, it is initially impossible to determine the specific multiplicity of a lensed GW  system upon the first detection. Therefore, we investigate the statistical properties of the entire population of multiple-image events. \autoref{fig:multi_snr_td} summarizes the joint distribution of the SNR ratio and time delay between subsequent and preceding signals for all lensed systems that exceed the detection threshold and contain at least two images. This dataset represents a more general population than the simple sum of double and quadruple systems, as it accounts for the broader detectability of lensed signals within a realistic observation window.

\begin{figure*}[t!]
    \centering
    \includegraphics[width=1\linewidth]{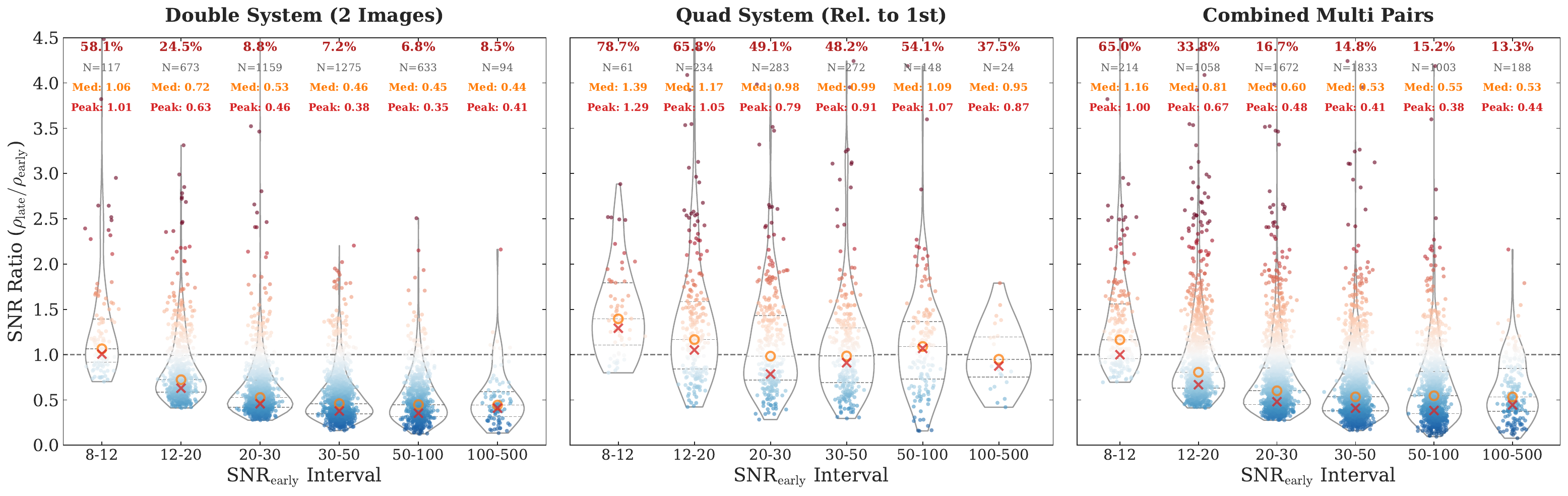}
    \caption{Evolution of the signal-to-noise ratio (SNR) ratio ($\rho_{\text{late}}/\rho_{\text{early}}$) distribution as a function of the early-arriving image's SNR, $\rho_{\text{early}}$. Violin plots visualize the probability density of the SNR ratios for double-image systems (left panel), quadruple-image systems (middle panel, relative to the first-arriving image), and combined multiple-image pairs (right panel) across six increasing SNR intervals. These intervals range from the near-threshold regime ($8\text{--}12$) to the high-loudness regime ($100\text{--}500$). Individual events are overlaid as scatter points, while trend lines track the evolution of the peak probability density. Annotations at the top of each bin indicate the percentage of events exhibiting a "flux ratio anomaly" ($\rho_{\text{late}} > \rho_{\text{early}}$, highlighted in red), the sample size ($N$), the median (Med), and the peak value (Peak). The horizontal dashed line marks the unity SNR ratio ($\rho_{\text{late}}/\rho_{\text{early}} = 1$).}
    \label{fig:snr_elect}
\end{figure*}
The global distribution remains dominated by double-image systems. Consequently, approximately $79.5\%$ of the events exhibit a late-arriving signal that is fainter than the early-arriving one, with a median SNR ratio of $0.62$ (and a peak at $0.47$). This confirms that the demagnification effect remains a prevalent feature across the lensed GW population.

In the time domain, the distribution of time delays shows a peak (mode) around $66.2$ days, with a median of $47.3$ days. However, the presence of a few extreme-delay events causes the distribution to be significantly right-skewed, resulting in a mean time delay of approximately $179.1$ days. This finding provides a crucial time baseline for future lensed GW observation missions: to ensure the capture of a complete set of multiple images, the detector network must maintain stable and continuous monitoring for at least six months.

Furthermore, correlation analysis indicates that the $r_p \approx -0.19$ and $r_s \approx -0.28$  are higher in magnitude compared to those in the double-image sub-population. This suggests an enhanced negative correlation between the SNR ratio and the time delay when considering the entire ensemble of lensed events.

To establish a prior basis for the identification of future lensed GW signals, we systematically investigated the joint distribution of the SNR ratio and the time delay. While  {foundational searches \citep[e.g.,][]{2018arXiv180707062H}} relied solely on the time delay $\Delta t$ as a prior,  {recent methodological advancements have demonstrated the necessity of a multi-dimensional approach. Notably, \citet{10.1093/mnras/stac1704} pioneered the inclusion of joint probability distributions—incorporating time delays, relative magnifications, and Morse phase shifts—to significantly improve the discrimination of lensed pairs from unlensed background events. More recently, \citet{Barsode_2025} highlighted that further improvements to this posterior overlap Bayes factor statistic (e.g., their PO2.0 framework) require rigorously incorporating informative population priors that account for measurement uncertainties and detector selection effects.}

 {Building upon these conceptual improvements,} we extend  {the traditional} framework by incorporating the SNR ratio $\rho_2/\rho_1$  {(which serves as a direct proxy for relative magnification) alongside the time delay}. Consequently, we propose  {an updated} Bayes factor formulation that accounts for both the time delay and the relative signal strength:
\begin{equation}
    \mathcal{R}^L_U = \frac{P(\Delta t_0, \rho_1,\rho_2/\rho_1 \mid \mathcal{H}_L)}{P(\Delta t_0, \rho_1, \rho_2/\rho_1 \mid \mathcal{H}_U)}.
    \label{eq:bayes_factor}
\end{equation}
If future observational data $(\Delta t_{\text{obs}}, q_{\text{obs}})$ fall within the high-density core region (indicated by deep blue in the figure), the value of the numerator $P(\Delta t_0, \rho_2/\rho_1 \mid \mathcal{H}_L)$ is maximized. This results in a significantly elevated Bayes factor $\mathcal{R}^L_U$, thereby providing strong support for the lensing hypothesis.  {Furthermore, as emphasized by \citet{Barsode_2025}, incorporating selection effects into these prior distributions is paramount for accurate identification. To explicitly account for} the impact of detector selection effects on the screening and authentication of lensed GW signals, we  {systematically} investigated the lensing characteristics across various SNR intervals.

We observe significant detector selection effects near the screening threshold. As illustrated in \autoref{fig:snr_elect}, a pronounced ``flux ratio anomaly'' is evident across all types of multiple-image events within the low-SNR interval ($\text{SNR} \in [8, 12]$).

This phenomenon arises from the intrinsic properties of typical double-image systems, where the second (late-arriving) image is generally less magnified than the first (early-arriving) image. Consequently, if the SNR of the first image ($\rho_{\mathrm{early}}$) is merely at the detector's threshold, the second image would likely fall below the detection limit. Therefore, the detection pipeline preferentially selects only those rare systems where the second image is significantly louder than the first (i.e., $\rho_{\mathrm{late}} > \rho_{\mathrm{early}}$). For instance, in double-image systems within the low-SNR interval, the proportion of events exhibiting this flux ratio anomaly reaches $58.1\%$, with a median ratio of $1.06$. This  {directly confirms} that evaluation criteria  {and prior distributions} must be adaptive across different SNR regimes.

Regarding quadruple systems (middle panel), the peak of the SNR ratio distribution is noticeably higher. This is attributed to the specific image configuration of quad-lens systems, where the first arriving image is not necessarily the brightest (e.g., it often corresponds to the third brightest image in the sequence). In the $\text{SNR}_{\text{early}} \in [8, 12]$ interval, the proportion of flux ratio anomalies in quadruple systems reaches $78.7\%$, with a median of $1.39$.

As the global SNR increases, the peak of the SNR ratio distribution consistently shifts toward lower values and eventually stabilizes. In double-image systems, the peak gradually decreases from $1.01$ and plateaus at approximately $0.35$. This trend  {clearly demonstrates that as the events become louder,} the impact of detector selection bias diminishes, allowing the statistical properties of the detected signals to converge toward their intrinsic physical distributions.

\begin{figure}[t!]
    \centering
    \includegraphics[width=1\linewidth]{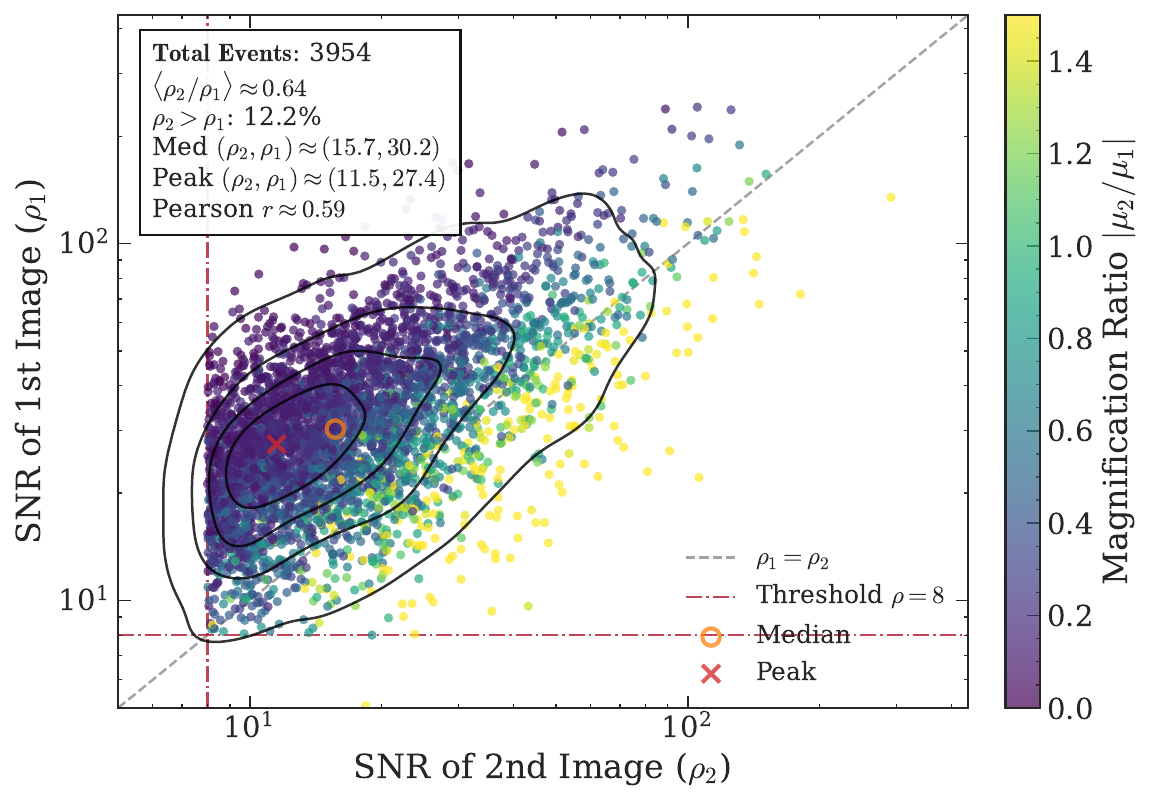}
    \caption{Joint SNR distribution of the first ($\rho_1$) and second ($\rho_2$) images for 3,954 double-image events. The distribution exhibits a moderate positive correlation (Pearson $r \approx 0.59$). The median of the distribution is located at $(\rho_2, \rho_1) \approx (15.7, 30.2)$, while the peak density occurs at $(11.5, 27.4)$. Approximately $12.2\%$ of events display ``flux inversion'' ($\rho_2 > \rho_1$), a phenomenon primarily attributed to the modulation of the detector response function by Earth's rotation and demagnification effects. The color gradient illustrates that deviations from the diagonal are principally governed by the magnification ratio $|\mu_2/\mu_1|$.}
    \label{fig:double_snr}
\end{figure}

\begin{figure*}[t!]
    \centering
    \includegraphics[width=1\textwidth]{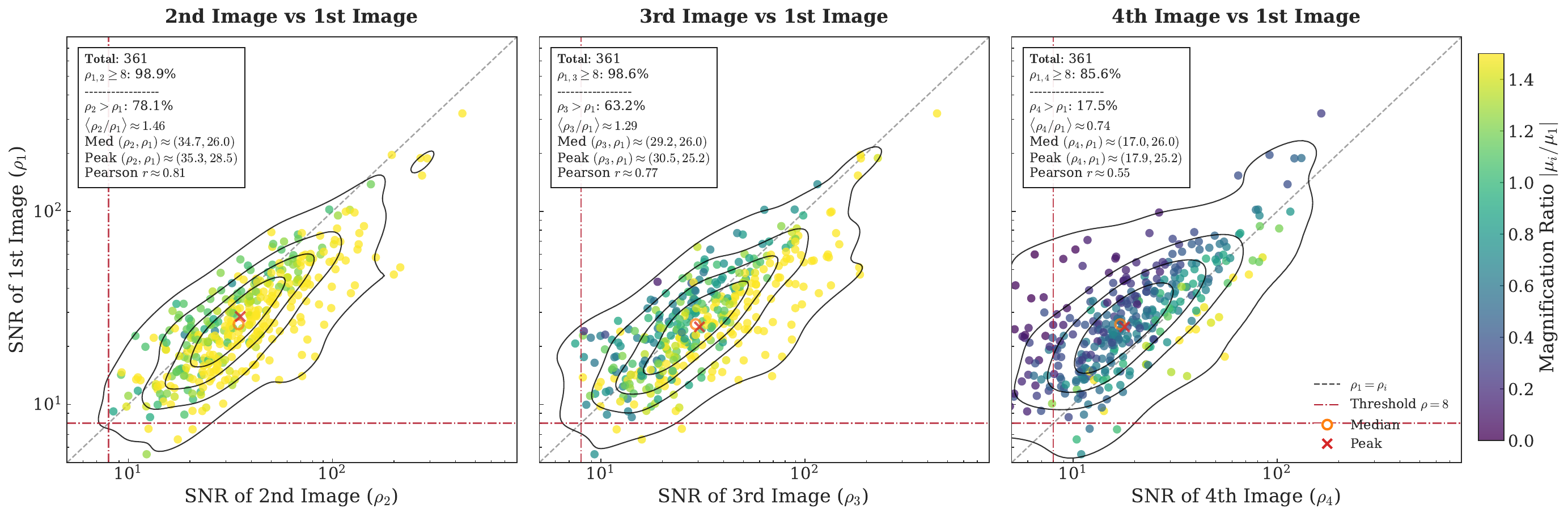}
    \caption{Joint SNR distributions between the first-arriving signal and the subsequent second, third, and fourth signals in 361 quadruple-lens systems. As the second and third images typically form near the lens caustics, they undergo significant magnification. Consequently, the earliest arriving signal ($\rho_1$) frequently ranks third in intensity. Furthermore, with $\rho_{1,4} \ge 8$ satisfied in only $85.6\%$ of cases, approximately $14.4\%$ of the fourth images fall below the detection threshold, rendering them unobservable.}
    \label{fig:quad_snr}
\end{figure*}

\autoref{fig:double_snr} illustrates the joint distribution of SNRs for the 3,954 analyzed double-image events. Statistical analysis reveals a Pearson correlation coefficient of $r \approx 0.59$ , indicating a moderate positive correlation between the SNRs of the first ($\rho_1$) and second ($\rho_2$) images. This degree of correlation is primarily constrained by the time-varying detector antenna response--driven by Earth's rotation--and the intrinsic diversity of lensing magnification ratios, which together act as significant decorrelation factors.The distribution's median is centered at $(\rho_2, \rho_1) \approx (15.7, 30.2)$ , while the peak density (mode) occurs at $(11.5, 27.4)$. These statistical benchmarks intuitively reflect the characteristic hierarchy in the observed sample, where the SNR of the first image generally exceeds that of the second. Furthermore, approximately $12.2\%$ of the events exhibit "flux inversion" ($\rho_2 > \rho_1$), a phenomenon largely attributed to the modulation of the antenna response function over the lensing time delay. This effect is particularly pronounced for image pairs with comparable magnifications formed in the vicinity of fold or cusp caustics.The color gradient in the figure further elucidates this relationship: data points near the diagonal ($\rho_1 = \rho_2$) correspond to image pairs with magnification ratios near unity, whereas the darker purple regions signify highly de-magnified pairs. This visually confirms that the observed SNR disparities are predominantly governed by the lensing magnification ratio rather than stochastic fluctuations in the intrinsic source parameters.

Quadruple-image systems typically originate from sources positioned near the lens caustics, resulting in the formation of a pair of highly magnified images. These dominant images predominantly appear as the second and third arrivals, whereas the earliest arriving signal typically ranks third in intensity---a phenomenon distinctly illustrated in \autoref{fig:quad_snr}. A sequential comparison between the first arrival and subsequent signals reveals a positive correlation between $\rho_1$ and $\rho_i$ (for $i=2,3,4$), with Pearson coefficients of 0.81, 0.77, and 0.55, respectively. The statistical medians further corroborate this intensity hierarchy: the medians for the second and third images against the first are $(\rho_2, \rho_1) \approx (34.7, 26.0)$ and $(\rho_3, \rho_1) \approx (29.2, 26.0)$ , with $78.1\%$ and $63.2\%$  of events exhibiting $\rho_2 > \rho_1$ and $\rho_3 > \rho_1$, respectively. Conversely, analysis of the $\rho_1$ versus $\rho_4$ subplot indicates that approximately $14.4\%$ of the fourth images fall below the detection threshold. Consequently, for quadruple-image candidates, prospective follow-up monitoring alone is insufficient; retrospective searches are crucial for recovering the complete lensing system.

\begin{figure}[t!]
    \centering
    \includegraphics[width=1\linewidth]{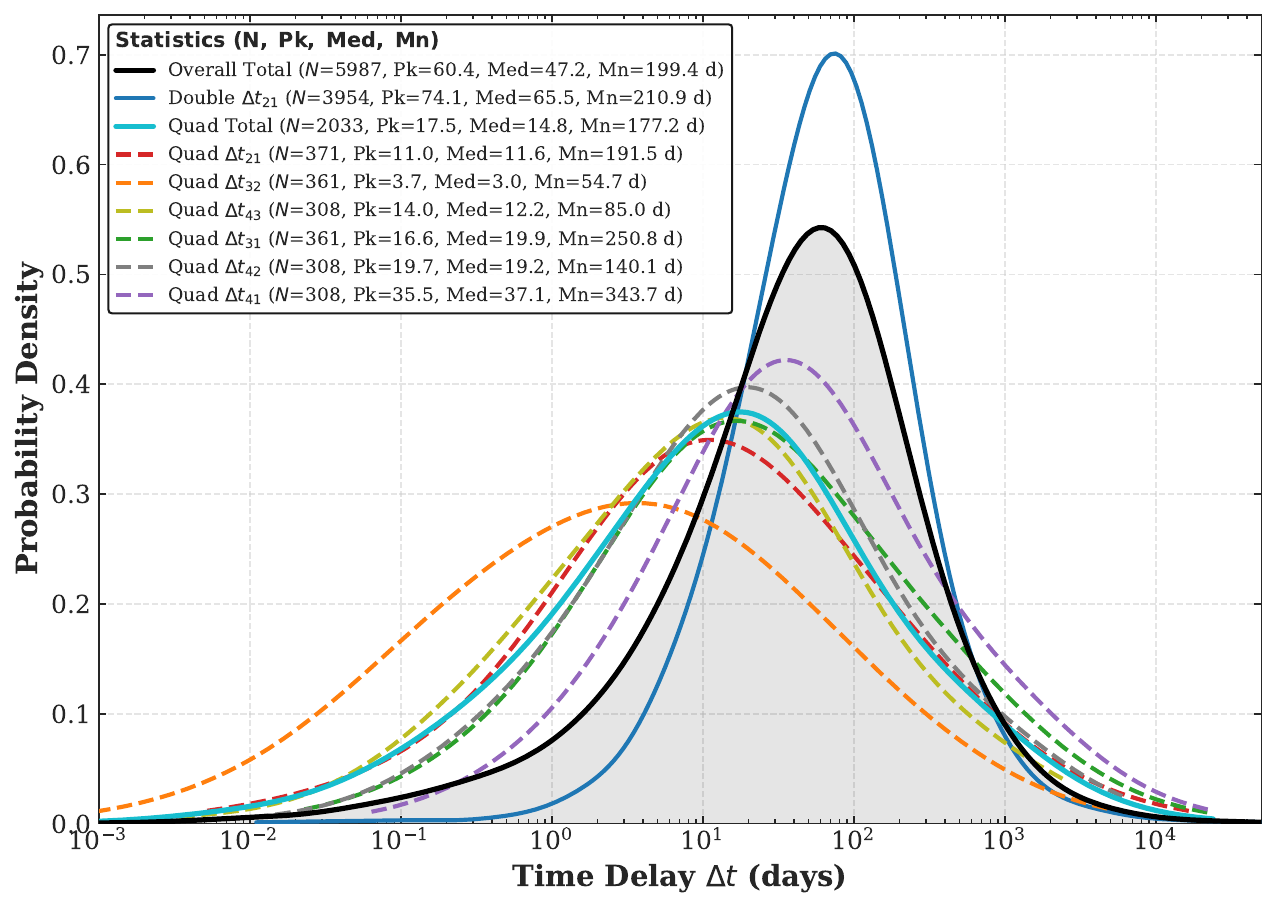}
    \caption{Probability density distributions of time delays for all multiple-image events within the $N_{\rm trig}$ population. This inclusive dataset encompasses quadruple systems where only two images are detectable; therefore, density curves for various image pair combinations are presented. Detailed specifications for each curve, including sample size ($N$), peak ($Pk$), median ($Med$), and mean ($Mn$), are provided in the legend.}
    \label{fig:td_density}
\end{figure}

To establish a time delay prior for identifying lensed GW pairs, we present the probability density distributions of time delays across various image pair combinations within multi-image systems, as illustrated in \autoref{fig:td_density}. Globally, the distributions for double systems, quadruple systems, and the aggregate population consistently exhibit a mean value exceeding the median, indicating a strong right-skewed characteristic. While the majority of events cluster around time delays of several tens of days, a subset of extreme events with delays spanning several years significantly elevates the overall mean.Focusing on double systems, the median time delay of 65.5 days represents a characteristic strong lensing timescale, reflecting the typical potential well depth generated by the mass distribution of lens galaxies. In quadruple systems, the time delays between different image pairs provide insight into the geometric properties near the lens caustics. The $\Delta t_{32}$ pair exhibits the shortest delay (median = 3.0 days) among all combinations; this typically corresponds to a pair of closely separated, highly magnified images with opposite parities, formed when the source crosses near a fold or cusp caustic. The intervals $\Delta t_{21}$ (median = 11.6 days) and $\Delta t_{43}$ (median = 12.2 days) represent the time differences between this "merging pair" (images 2 and 3) and the earlier first image or the later fourth image, respectively. Meanwhile, $\Delta t_{41}$ (median = 37.1 days, mean = 343.7 days) characterizes the full system span, indicating the total duration of the quadruple lensing event.

Consequently, if detectors observe two signals with highly consistent parameters within a few days, they likely correspond to the high-magnification pair of a quadruple system. To capture the complete signal set of such a system, attention must be directed towards identifying the potential first signal that arrived earlier and the fourth signal that may arrive subsequently. Governed by the overall time delay distribution, we assign a higher prior probability of being a lensed GW to signal pairs with time delays near the population median of approximately 47.2 days.

\section{Conclusions and discussions}
In this work, we have constructed a comprehensive mock catalog of strongly lensed GW  events, employing a composite lens mass model that incorporates both dark matter halos and baryonic galaxy components.
While most of the current event rate calculations were based on SIS/SIE lens model, we adopted a more realistic halo model approach according to \cite{Abe_2025} who considered lenses modeled with five components: host halos, subhalos, galaxies, satellites and shear. Different from optical lenses which are limited by the spatial resolution, GW lenses can resolve images in time domain very well, thus subhalos which generate images with small time delays are a very important ingredient not considered in previous GW lensing studies. Strong lensing time delay necessitates proper treatment of the Earth rotation, which affects the antenna pattern dependent part of the detected wave strain. Therefore, GW signals from multiple images will differ from each other not only due to magnification, but also due to Earth rotation. We considered this effect carefully.

We presented predictions for lensed event rates across three distinct source populations---binary black holes (BBHs), binary neutron stars (BNSs), and neutron star-black hole binaries (NSBHs)---under four different detector network configurations: A+, Cosmic Explorer (CE), Einstein Telescope (ET), and a combined CE+ET network. Our simulations assume Planck18 cosmology and a low-metallicity standard isolated binary evolution model \cite{2013ApJ...779...72D}. Regarding the source population settings, we integrated redshift distributions consistent with GWTC-4 observations \cite{2025arXiv250818082T} and adopted specific mass distribution models: GWTC-4 models for BBH and BNS systems, and the model from \cite{10.1093/mnras/stac3052} for NSBH systems. We utilized rigorous statistical sampling techniques, such as rejection sampling and inverse transform sampling, to generate specific source parameters. To unify the treatment of lensing effects across different source bands, we constructed an equivalent luminosity function based on the GW redshift and SNR distributions, enabling the efficient generation of lensed event sequences.

Building upon this framework, we provided a detailed and complete set of predictions for lensed multi-image systems, encompassing both the standard multi-image systems that form the bulk of the catalog and rare events with unique physical significance. We focused our analysis on the following categories:

\begin{itemize}
    \item \textbf{Standard Multi-Image Systems (Doubles and Quads):} Representing the most typical image configurations resulting from strong lensing, these systems constitute the majority of our mock catalog. We presented complete parameter distributions for these systems under various detector networks, including time delays between images, relative SNR ratios, and angular separations. These fundamental statistical data quantify the observable characteristics of typical lensed events, providing core statistical priors for the development of future lensing search strategies and the evaluation of detection efficiencies.

    \item \textbf{Events lensed by Subhalos:} Within the multi-image systems, approximately 25\% of events exhibit significant perturbations due to subhalos (manifesting as wave diffraction or interference effects). These events serve as crucial samples for investigating dark matter substructures within lens galaxies.

    \item \textbf{Central Image Events:} We identified a class of central image events distinct from those in optical lensing systems. Thanks to the extreme sensitivity of third-generation detectors (such as CE and ET), central images---which are typically unobservable in the optical band due to severe demagnification---are expected to be detectable in the GW band. This discovery provides a key probe for testing the odd-image theorem of gravitational lensing and exploring the physical properties of lens galaxy centers, such as the core mass distribution.

    \item \textbf{High-Magnification Events:} We recorded all high-magnification events where the maximum magnification in any lens system exceeds 3 or 10. Due to the magnification bias effect, these events allow us to detect GW signals from higher redshifts, holding significant value for understanding the formation and evolution of the early universe. Among lensed supernovae, quasars and GWs, the ratios of such highly magnified events sequentially decrease since their luminosities/``equivalent luminosities" sequentially increase. For brither sources, the geometry (source, lens and the earth are on a line) are more important while selection effect becomes more important for fainter sources.

    \item \textbf{Small-time-delay Events:} Different from optical case which is limited by spatial resolution, a small part of lensed GWs could exist with small time delays since the time-domain resolution is quite high due to the transient nature of GWs in ground-based cases.
\end{itemize}

Furthermore, we screened all lensed GW events that crossed the detector thresholds (applying a looser criterion compared to that for multi-image systems) and performed a comparative analysis of model sensitivity. Our findings indicate that the lensed event rate is highly sensitive to the binary evolution model, depending directly on the total number of GW mergers predicted by different models. Regarding the mass model, the Truncated Power Law + 2 Peak model proved robust across different mass ranges, whereas the Power Law model exhibited higher sensitivity.

Finally, we provided a series of detailed distributions for the simulated results, including the joint distribution of source and lens redshifts, the distribution of maximum angular separations, and the distributions of inter-image SNR ratios and time delays. We also quantified the impact of selection effects. These results offer precise Bayesian priors for the future authentication of lensed GW signals. All data products generated in this study, along with the complete Mock Catalog, have been publicly released on GitHub under the project name \texttt{GW-LMC} (https://github.com/LensedGW/GW-LMC) for community use.

Due to the length limit, we have only discussed the systems that can be detected in this paper. Meanwhile, a companion work studying the events that can be identified basing on the catalogs we have built is being prepared as well as a work focusing on space-borne detector case.

\section*{Acknowledgments}
We thank Masamune Oguri for helping in using SLHammocks. This work was supported by National Key R$\&$D Program of China (No. 2024YFC2207400). X.Ding was supported by National Natural Science Foundation of China (Grant No. 12573017). M.B. was supported by the Polish National Science Centre Grants: $2023/50/A/ST9/00579$ and $ 2023/49/B/ST9/02777$. M.B. gratefully acknowledges the support from COST Action CA21136 - ``Addressing observational tensions in cosmology with systematics and fundamental physics (CosmoVerse)".

\section*{Data Availability}
The mock catalogs generated in this work are available in the GW-LMC repository on GitHub: \url{https://github.com/LensedGW/GW-LMC} and have been archived on Zenodo \url{https://doi.org/10.5281/zenodo.19212271}.

\bibliography{myrefs}{}
\bibliographystyle{aasjournal}
\end{document}